# THE EVOLUTION OF THE GALAXY STELLAR MASS FUNCTION AT $z = 4$–8: A STEEPENING LOW-MASS-END SLOPE WITH INCREASING REDSHIFT

Mimi Song[1], Steven L. Finkelstein[1], Matthew L. N. Ashby[2], A. Grazian[3], Yu Lu[4], Casey Papovich[5], Brett Salmon[5], Rachel S. Somerville[6], Mark Dickinson[7], K. Duncan[8,9], Sandy M. Faber[10], Giovanni G. Fazio[2], Henry C. Ferguson[11], Adriano Fontana[3], Yicheng Guo[10], Nimish Hathi[12], Seong-Kook Lee[13], Emiliano Merlin[3], S. P. Willner[2]

*Accepted by ApJ*

## ABSTRACT

We present galaxy stellar mass functions (GSMFs) at $z = 4$–8 from a rest-frame ultraviolet (UV) selected sample of $\sim$4500 galaxies, found via photometric redshifts over an area of $\sim$280 arcmin$^2$ in the CANDELS/GOODS fields and the Hubble Ultra Deep Field. The deepest *Spitzer*/IRAC data yet-to-date and the relatively large volume allow us to place a better constraint at both the low- and high-mass ends of the GSMFs compared to previous space-based studies from pre-CANDELS observations. Supplemented by a stacking analysis, we find a linear correlation between the rest-frame UV absolute magnitude at 1500 Å ($M_{\rm UV}$) and logarithmic stellar mass ($\log M_*$) that holds for galaxies with $\log(M_*/M_\odot) \lesssim 10$. We use simulations to validate our method of measuring the slope of the $\log M_*$–$M_{\rm UV}$ relation, finding that the bias is minimized with a hybrid technique combining photometry of individual bright galaxies with stacked photometry for faint galaxies. The resultant measured slopes do not significantly evolve over $z = 4$–8, while the normalization of the trend exhibits a weak evolution toward higher mass at higher redshift. We combine the $\log M_*$–$M_{\rm UV}$ distribution with observed rest-frame UV luminosity functions at each redshift to derive the GSMFs, finding that the low-mass-end slope becomes steeper with increasing redshift from $\alpha = -1.55^{+0.08}_{-0.07}$ at $z = 4$ to $\alpha = -2.25^{+0.72}_{-0.35}$ at $z = 8$. The inferred stellar mass density, when integrated over $M_* = 10^8$–$10^{13}$ $M_\odot$, increases by a factor of $10^{+30}_{-2}$ between $z = 7$ and $z = 4$ and is in good agreement with the time integral of the cosmic star formation rate density.

*Subject headings:* galaxies: evolution — galaxies: formation — galaxies: high-redshift — galaxies: mass function

## 1. INTRODUCTION

The near-infrared (near-IR) capability of the Wide Field Camera 3 (WFC3) on board the *Hubble Space Telescope* (*HST*) has now yielded a statistically significant sample of galaxies in the early universe, enabling

mmsong@astro.as.utexas.edu
[1] Department of Astronomy, The University of Texas at Austin, C1400, Austin, TX 78712, USA
[2] Harvard-Smithsonian Center for Astrophysics, 60 Garden St., Cambridge, MA 02138, USA
[3] INAF – Osservatorio Astronomico di Roma, via Frascati 33, 00040 Monteporzio, Italy
[4] Observatories, Carnegie Institution for Science, Pasadena, CA, 91101
[5] Department of Physics and Astronomy, Texas A&M University, College Station, TX 77843, USA
[6] Department of Physics & Astronomy, Rutgers University, 136 Frelinghuysen Road, Piscataway, NJ 08854
[7] National Optical Astronomy Observatory, Tucson, AZ 85719
[8] University of Nottingham, School of Physics & Astronomy, Nottingham NG7 2RD
[9] Leiden Observatory, Leiden University, NL-2300 RA Leiden, The Netherlands
[10] University of California Observatories/Lick Observatory, University of California, Santa Cruz, CA, 95064
[11] Space Telescope Science Institute, Baltimore, MD 21218
[12] Aix Marseille Universite, CNRS, LAM (Laboratoire d'Astrophysique de Marseille) UMR 7326, 13388, Marseille, France
[13] Center for the Exploration of the Origin of the Universe (CEOU), Astronomy Program, Department of Physics and Astronomy, Seoul National University, Shillim-Dong, Kwanak-Gu, Seoul 151-742, Korea

us to pass the era of simply discovering very distant galaxies and enter an era where we can perform systematic studies to probe the underlying physical processes. Such studies have begun to make progress toward understanding galaxy evolution at high redshift, with particular advances in measurements of the rest-frame ultraviolet (UV) luminosity function (e.g., Bouwens et al. 2011; Oesch et al. 2012; Schenker et al. 2013b; Lorenzoni et al. 2013; Finkelstein et al. 2015; Bouwens et al. 2015), the rest-frame UV spectral slope (e.g., Finkelstein et al. 2012; Bouwens et al. 2014), and the cosmic star formation rate density (SFRD; see Madau & Dickinson 2014 for a review).

A key constraint on galaxy evolution that has only recently begun to be robustly explored is that of the growth of stellar mass in the universe. This measurement requires a combination of deep rest-frame UV data with constraints at rest-frame optical wavelengths, to better probe the emission from older, lower-mass stars. The mass assembly history across cosmic time is governed by complicated processes, including star formation, merging of galaxies, supernova feedback, etc. In spite of the complexity of the baryonic physics of galaxy formation, however, various studies have found that global galaxy properties, such as star formation rate (SFR), metallicity, size, etc., all correlate tightly with the stellar mass (Kauffmann et al. 2003b; Noeske et al. 2007; Tremonti et al. 2004; Williams et al. 2010), implying that the stellar mass plays a major role in galaxy evolution. Thus, de-termining the comoving number density of galaxies in a wide range of stellar masses (i.e., the galaxy stellar mass function [GSMF]) and following the evolution with redshift constitutes a basic and crucial constraint on galaxy formation models. Specifically, obtaining robust observational constraints on the low-mass-end slope of the GSMF can provide insights on the impact of feedback on stellar mass buildup of low-mass galaxies (Lu et al. 2014), as current theoretical models predict steeper low-mass-end slopes than those previously observed (e.g., Vogelsberger et al. 2013; see Somerville & Davé 2014 for a review). Consequently, the evolution of GSMFs over the past 11 billion years has been extensively investigated observationally during the past decade (e.g., Marchesini et al. 2009; Baldry et al. 2012; Ilbert et al. 2013; Muzzin et al. 2013; Tomczak et al. 2014), mostly using traditional techniques (e.g., $1/V_{max}$, maximal likelihood) that assess and correct for the incompleteness in mass.

Nonetheless, the GSMF in the first 2 billion years after the Big Bang has remained poorly constrained, due to (i) limited sample sizes, particularly of low-mass galaxies at high redshift; (ii) systematic uncertainties in stellar mass estimations; and (iii) the lack of *Spitzer Space Telescope* Infrared Array Camera (IRAC; Fazio et al. 2004) data with comparable depth to *HST*/WFC3. IRAC data are essential to probe stellar masses of galaxies at $z \gtrsim 4$ as the 4000 Å/Balmer break moves into the mid-infrared (mid-IR), beyond the reach of *HST* (and ground-based telescopes).

An alternative approach to deriving the GSMF takes advantage of the fact that the selection effects and incompleteness are relatively well known and corrected for in rest-frame UV luminosity functions. Therefore, one can derive GSMFs by taking a UV luminosity function and convolving it with a stellar mass versus UV luminosity distribution at each redshift. Using this approach, González et al. (2011) derived GSMFs at $4 < z < 7$ of Lyman break galaxies (LBGs) over $\sim$33 arcmin$^2$ of the Early Release Science (ERS) field (for $z = 4$–6, and the Hubble Ultra Deep Field [HUDF] and the Great Observatories Origins Deep Survey [GOODS] fields with pre-WFC3 data for $z = 7$) utilizing the *HST*/WFC3 and IRAC GOODS-South data. They reported a shallow low-mass-end slope of $-(1.4–1.6)$ at $z = 4$–7, but the small sample size and limited dynamic range made it difficult for them to explore the mass-to-light distribution at $z > 4$, and their GSMFs were derived under an assumption that the mass-to-light distribution at $z \sim 4$ is valid up to $z \sim 7$.

More recently, several studies have utilized the large *HST* data set from the Cosmic Assembly Near-infrared Deep Extragalactic Legacy Survey (CANDELS; Grogin et al. 2011; Koekemoer et al. 2011) to make progress on the measurements of the GSMFs at $z > 4$. Duncan et al. (2014) and Grazian et al. (2015) derived the GSMFs of galaxies at $4 < z < 7$ in the GOODS-S field and GOODS-S/UDS fields, respectively, using the CANDELS *HST* data and the *Spitzer* Extended Deep Survey (SEDS; Ashby et al. 2013) IRAC data. These studies have reported a steeper low-mass-end slope of $\alpha \sim -(1.6–2.0)$ than previous studies, but the uncertainties are still large, and the inferred evolution of the low-mass-end slope of the GSMFs remains uncertain.

Here we probe galaxy buildup from $z = 4$ out to $z = 8$, aiming to improve on the limiting factors discussed above, with a goal of providing robust constraints on the GSMFs of galaxies at $4 < z < 8$. We do this by combining near-IR data from CANDELS GOODS-S and GOODS-N fields with the deepest existing IRAC data over the GOODS fields from the *Spitzer*-CANDELS (S-CANDELS; PI Fazio; Ashby et al. 2015) and the IRAC Ultra Deep Field 2010 Survey (UDF10; Labbé et al. 2013). We obtain reliable photometry on these deep IRAC data by performing point-spread-function (PSF) matched deblending photometry, enabling us to extend the exploration of the GSMFs to lower stellar masses and higher redshifts than previous studies. Furthermore, a special emphasis is put on quantifying and minimizing the systematics inherent in our analysis via mock galaxy simulations. While taking a similar approach of convolving a rest-frame UV luminosity function with stellar mass to rest-frame UV luminosity distribution as González et al. (2011), the increased sample size and deeper data enable us to bypass the limitations of the previous studies, as we measure the mass-to-light ratio distribution at every redshift.

This paper is organized as follows. Section 2 introduces the *HST* data sets used in this study, as well as our sample at $3.5 < z < 8.5$ selected by photometric redshifts, and discusses our IRAC photometry, which is critical for the stellar mass estimation described in Section 3. Section 4 presents stellar mass versus observed rest-frame absolute UV magnitude ($M_*$–$M_{UV}$) distributions and introduces a stacking analysis and mock galaxy simulations. By combining the rest-frame UV luminosity function with the $M_*$–$M_{UV}$ distribution, we derive GSMFs and stellar mass densities in Section 5 and 6, respectively. A discussion and summary of our results follow in Section 7 and Section 8. Throughout the paper, we use the AB magnitude system (Oke & Gunn 1983) and a Salpeter (1955) initial mass function (IMF) between 0.1 $M_\odot$ and 100 $M_\odot$. All quoted uncertainties represent 68% confidence intervals unless otherwise specified. We adopt a concordance $\Lambda$CDM cosmology with $H_0 = 70 = 100h$ km s$^{-1}$ Mpc$^{-1}$, $\Omega_M = 0.3$, and $\Omega_\Lambda = 0.7$. The *HST* bands F435W, F606W, F775W, F814W, F850LP, F098M, F105W, F125W, F140W, and F160W will be referred to as $B$, $V$, $i$, $I_{814}$, $z$, $Y_{098}$, $Y$, $J$, $JH_{140}$, and $H$, respectively.

## 2. DATA

Constraining GSMFs requires a deep multiwavelength data set over a wide area in order to probe the full dynamic range of a galaxy population. In this section, we describe the *HST* imaging used to select our galaxy candidates, as well as the candidate selection process. We then discuss the procedures used to measure accurate photometry for these galaxies from the *Spitzer*/IRAC S-CANDELS imaging.

### 2.1. HST Data and Sample Selection

The galaxy sample employed in this study is from Finkelstein et al. (2015), to which we refer the reader for full details of the *HST* data used and the galaxy sample selection. This sample consists of $\sim$7000 galaxies selected via photometric redshifts over a redshift range of $z = 3.5$–8.5. These galaxies were selected using *HST* imaging data from the CANDELS (Grogin et al. 2011; Koekemoer et al. 2011) over the GOODS (Giavalisco et al. 2004) North (GOODS-N) and South (GOODS-S) fields, the ERS (Windhorst et al. 2011) field, and the HUDF (Beckwith et al. 2006; Bouwens et al. 2010; Ellis et al. 2013) and its two parallel fields (Oesch et al. 2007; Bouwens et al. 2011).[14] We use the full data set which incorporates all earlier imaging from *HST* Advanced Camera for Surveys (ACS), including the *B*, *V*, *i*, $I_{814}$, and *z* filters. We also use imaging from the *HST* WFC3 in the $Y_{098}$, *Y*, *J*, $JH_{140}$, and *H* filters. A complete description of these data is presented by Koekemoer et al. (2011, 2013). These combined data are three-layered in depth, comprising the extremely deep HUDF, moderately deep CANDELS-Deep fields, and relatively shallow CANDELS-Wide and ERS fields, designed to efficiently sample both bright and faint galaxies at high redshifts.

| F160W | 3.6 $\mu$m | 3.6 $\mu$m Model | 3.6 $\mu$m Residual |
|---|---|---|---|

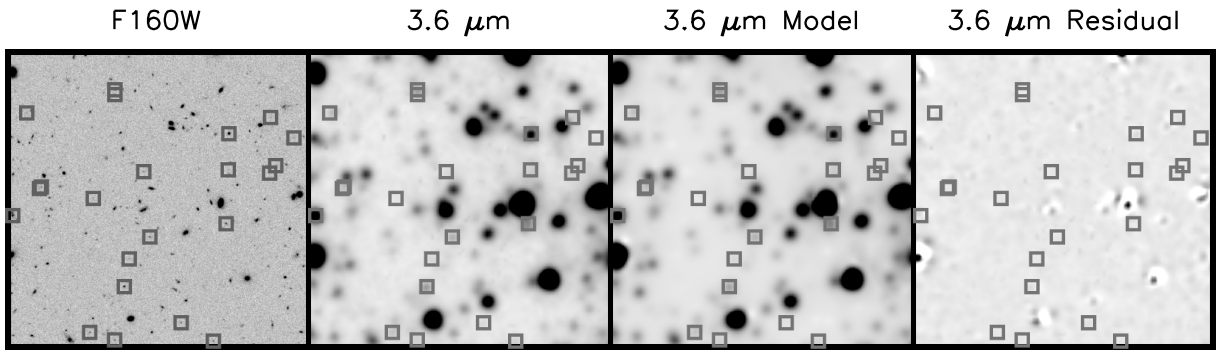

FIG. 1.— Example of our IRAC photometry modeling procedure. From *left* to *right*: (1) *H*-band WFC3 imaging of a $1' \times 1'$ region in the GOODS-S field, (2) S-CANDELS 3.6 $\mu$m imaging, (3) the best-fit 3.6 $\mu$m model image, and (4) the T-PHOT residual image (i.e., real science image subtracted by the best-fit model image). Our high-redshift galaxies (gray squares) are often blended with nearby foreground sources; therefore, we perform PSF-matched photometry on the S-CANDELS data using the WFC3 *H*-band imaging as a prior on the position and morphology of sources. Using the T-PHOT software package, we convolve the *H*-band image with empirically derived IRAC PSFs to generate low-resolution (IRAC) model images, allowing the flux of each source to vary to simultaneously fit all sources in the IRAC data.

As described by Finkelstein et al. (2015), sources were detected in a summed $J + H$ image, using a more aggressive detection scheme compared to that used to build the official CANDELS catalog (e.g., Guo et al. 2013, Barro et al. 2016, in preparation) to detect faint sources at high redshifts, yielding a catalog with the total number of sources a factor of two larger.[15] To minimize the presence of spurious sources, only those objects with $\geq 3.5\sigma$ significance in both the *J* and *H* bands were evaluated as possible high-redshift galaxies. Photometric redshifts were estimated with EAZY (Brammer et al. 2008), and selection criteria devised based on the full redshift probability density function (pdf; $P(z)$) were applied (Finkelstein et al. 2015). These criteria were designed to assess the robustness of the sample, requiring the primary redshift solution to have more than 70% of the integrated redshift pdf, and the narrowness of $P(z)$, by limiting the sample to those for which the integral of the redshift pdf for the corresponding redshift bin is at least 25% of the total integral of the pdf. A comparison of our photometric redshifts for 171 galaxies with spectroscopic redshifts in our $z > 3.5$ sample shows an excellent agreement, with $\sigma(\Delta z/(1 + z_{\rm spec})) = 0.03$ (after $3\sigma$ clipping). All of the selected sources were visually inspected for removal of artifacts and stellar contaminants. Active galactic nuclei (AGNs) identified in X-rays were also excluded from the sample. Our final parent sample consists of 4156, 2056, 669, 284, and 77 galaxies at $z = 4$ ($3.5 \leq z < 4.5$), 5 ($4.5 \leq z < 5.5$), 6 ($5.5 \leq z < 6.5$), 7 ($6.5 \leq z < 7.5$), and 8 ($7.5 \leq z < 8.5$), respectively.

### 2.2. *IRAC Data and Photometry*

At $3.5 < z < 8.5$, the observed mid-IR probes rest-frame optical wavelengths redward of the Balmer/4000 Å break. Deep *Spitzer*/IRAC data are therefore critical to constrain stellar masses and the resulting GSMF. One of the key advances of our study is the significantly increased depth in the 3.6 and 4.5 $\mu$m IRAC bands provided by the new S-CANDELS survey (Ashby et al. 2015). The final S-CANDELS mosaics in the GOODS-S and GOODS-N fields (where the former includes the HUDF parallel fields) include data from three previous studies: GOODS, with integration time of 23–46 hr per pointing (Dickinson et al., in preparation); a $5' \times 5'$ region in the ERS observed to 100 hr depth (PI Fazio); and the IRAC Ultra Deep Field 2010 program, which observed the HUDF and its two parallel fields to 120, 50, and 100 hr, respectively (Labbé et al. 2013). The total integration time within the S-CANDELS fields more than doubles the integration time for most of the area used in this study (to $\geq$50 hr total), reaching a total formal depth of 26.5 mag ($3\sigma$) at 3.6 $\mu$m and 4.5 $\mu$m (Ashby et al. 2015). Imaging at 5.8 and 8.0 $\mu$m was obtained with IRAC as part of the GOODS program. However, these data are too shallow to provide meaningful constraints for high-redshift faint galaxies, so we do not include them in our analysis.

The full-width at half-maximum (FWHM) of the PSF of the IRAC data ($\sim 1''.7$ at 3.6 $\mu$m versus $\sim 0''.19$ at 1.6 $\mu$m with WFC3) results in non-negligable source confusion, making accurate flux determinations challenging, as shown in Figure 1. The second panel of Figure 1 shows that our high-redshift galaxies are extremely faint and are often blended with nearby bright sources in the mid-IR, making simple aperture photometry unreliable. For reliable IRAC photometry on the deep S-CANDELS data, we therefore perform PSF-matched photometry using the T-PHOT software (Merlin et al. 2015), an updated version of TFIT (Laidler et al. 2007), on the S-CANDELS 3.6 and 4.5 $\mu$m mosaics. This PSF-matched photometry uses information in a high-resolution image (here the $H$ band), such as position and morphology, as priors. Specifically, we use isophotes and light profiles from the detection ($J+H$) image obtained by the Source Extractor package (SExtractor; Bertin & Arnouts 1996). The high-resolution image was convolved with a transfer kernel to generate model images for the low-resolution data (here the IRAC imaging), allowing the flux in each source to vary. This model image was in turn fitted to the real low-resolution image. The IRAC fluxes of sources are determined by the model that best represents the real data.

As the PSF FWHM of the high-resolution image ($H$ band) is negligible ($\sim 0\rlap{.}''19$) when compared to those of the low-resolution IRAC images ($\sim 1\rlap{.}''7$), we used IRAC PSFs as transfer kernels. We derive empirical PSFs in each band and in each field by stacking isolated and moderately bright ([3.6], [4.5] = 15.5–19.0 mag) stars identified in a half-light radius versus magnitude diagram in IRAC imaging. As T-PHOT requires the kernel to be in the pixel scale of the high-resolution image, each star image was 10 times oversampled to generate the final PSFs in the same pixel scale of the $H$-band image ($0\rlap{.}''06$ pixel$^{-1}$). They were then registered, normalized, and median-combined to generate the final IRAC PSFs.

Preparatory work for running T-PHOT includes performing background subtraction on the S-CANDELS mosaics using the script subtract_bkgd.py (written by H. Ferguson; see Galametz et al. 2013 for details) and dilation of the $J+H$ SExtractor segmentation map, analogous to the traditional aperture correction. The need for this dilation step originates from the fact that under nonzero background fluctuations, isophotes of faint sources include a smaller fraction of their total flux compared to bright sources, and therefore their IRAC fluxes based on isophotes tend to be underestimated. To include the light in the wings and to counterbalance the underestimation of flux for faint sources of which the amount depends on the isophotal area, an empirical correction factor to enlarge the SExtractor segmentation map was devised by the CANDELS team by comparing isophotal area from deep and shallow images (see Galametz et al. 2013 and Guo et al. 2013 for details). We applied this empirical correction factor to the $J+H$ segmentaion map using the program dilate (De Santis et al. 2007) while preventing merging between sources.

To correct for potential small spatial distortions or mis-registrations between the high-resolution and low-resolution images, a second run of T-PHOT was performed using a shifted kernel built by cross-correlating the model and real low-resolution images. Figure 1 presents an example of our IRAC photometry procedure on a subregion in GOODS-S. With the exception of very bright sources, the residual image is remarkably clean, highlighting the accuracy of this procedure.

### 2.2.1. *Verification*

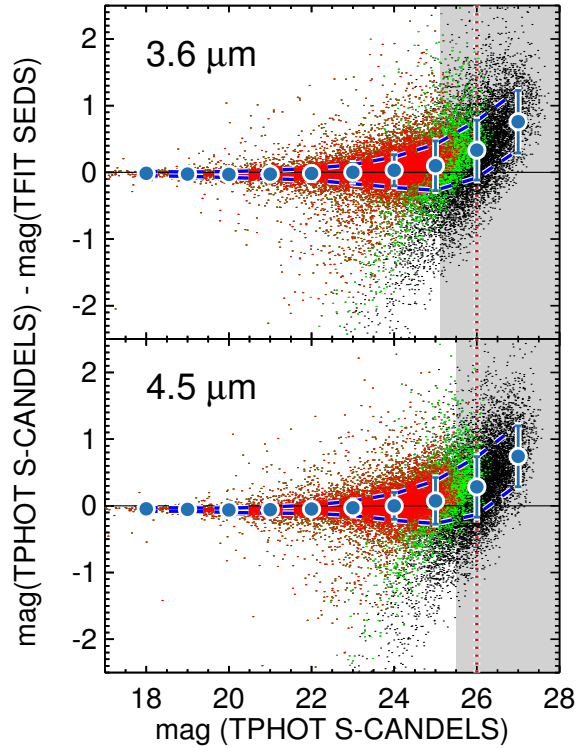

Fig. 2.— Comparison between our T-PHOT S-CANDELS photometry and the official CANDELS TFIT SEDS photometry from Guo et al. (2013) and Barro et al. (2016, in preparation) for IRAC 3.6 $\mu$m (*upper*) and 4.5 $\mu$m (*lower*) bands for sources with S/N > 1 (black), S/N > 3 (green), and S/N > 5 (red) in both catalogs. Blue circles and error bars indicate the median and robust standard deviation of the magnitude difference in each magnitude bin for sources with S/N > 1, and the blue dashed lines encompass the central 68% of the distribution. The red vertical dotted line denotes the $5\sigma$ S-CANDELS depth. The bias seen at faint magnitudes is due to Eddington bias of upscattered sources in the shallower SEDS data, which is shown as the gray shaded area (for the S/N > 1 cut; see Section 2.2.1 for more details). The SEDS data are shallower; thus, the agreement between the two catalogs in magnitude bins brighter than this range indicates that our photometry is accurate.

We tested the accuracy of our IRAC photometry in two ways. First, we compared our catalog with the official CANDELS catalogs (Guo et al. 2013; Barro et al. 2016, in preparation) in which IRAC fluxes were obtained from the shallower SEDS data using TFIT. Figure 2 shows the comparison between our magnitude and that from the official catalog for sources with signal-to-noise ratio (S/N) greater than 1, 3, and 5 in both catalogs. As we compare two catalogs generated from images with different depths with a certain S/N cut, faint sources are dominated by background fluctuations and the Eddington bias of upscattered sources in the shallower SEDS data, making the comparison unreliable (see Figure 13 of Guo et al. for a similar trend and discussion of the flux comparison). Taking a similar approach to that of Guo et al. (2013), we estimated this magnitude range in which the flux comparison is unreliable due to the Eddington bias for the S/N > 1 cut. We first found the best-fit power law to the differential number count density of the sources in the official CANDELS catalog without any S/N cut. Then, we compared the differential number count density of sources with S/N > 1 with the best-fit power law to find the magnitude that the former starts falling below 80% of what is expected by the best-fit power law ($[3.6] = 25.1$ and $[4.5] = 25.5$; gray shaded region in Figure 2). In magnitude bins brighter than this range, the comparison indicates an excellent agreement with a negligible systematic offset.

Second, we performed a mock source simulation in order to validate our photometry. Briefly, mock sources with varying physical properties (e.g., size, light profile, luminosity, redshift) were generated using the GALFIT software (Peng et al. 2002), of which flux densities in each band are assigned using the updated Bruzual and Charlot (CB07; Bruzual & Charlot 2003) stellar population synthesis (SPS) models. While the exact shape of the assumed distribution of physical parameters such as size or light profile can impact the detection rate of sources close to the sensitivity limit and thus the results of simulations designed to correct for completeness, T-PHOT photometry is by design limited to the sources recovered in the high-resolution detection image. Therefore, our simulation results should not be sensitive to those assumptions in the first order. These mock sources were convolved with the PSF of each filter and added at random locations in the ($H$-band PSF-convolved) high-resolution and IRAC low-resolution images. Our simulation thus accounts fully for source confusion with nearby foreground real sources, but we constrain the number density of mock sources to be negligible (5 arcmin$^{-2}$) compared to that of real sources to ensure that our simulation results are not dominated by self-crowding among the inserted artificial sources. The mock sources were then recovered using the same procedures as for real sources, including T-PHOT photometry. Figure 3 shows a comparison between the input and recovered 3.6 $\mu$m and 4.5 $\mu$m magnitudes. It is encouraging that we find no systematic offset between the two down to the 50% completeness limit of 25 mag (Ashby et al. 2015), given that the IRAC photometry is affected with many sources of uncertainty that demand accurate background subtraction, aperture correction scheme (dilation), and buildup of transfer kernels. As any observed offsets from these simulations are not statistically significant, we do not apply any correction to our observed IRAC photometry.

### 2.2.2. Visual inspection

A practical limitation exists when building empirical IRAC PSFs for deep fields, as such surveys by design target a relatively small area well off the Galactic plane, resulting in few stars present in the imaging. The severe source confusion in the IRAC data further reduces the number of isolated stars that can be used for the creation of PSFs. Therefore, although our PSF-matched IRAC photometry significantly improves photometric accuracy over more traditional methods (Lee et al. 2012), our IRAC photometry may be imperfect. Its significance, however, is likely small, as already discussed in the previous sections.

Another source of uncertainty is that TFIT/T-PHOT assumes no morphological $k$-corrections (no variation in the surface profile or morphology) between short- and long-wavelength images, which is likely not the case (although we attempted to mitigate this by using the reddest *HST* image as our high-resolution image). However, our high-redshift galaxies are small enough to not be resolved in the low-resolution image (see Figure 25 of Ashby et al. 2013), and thus it should have a negligible effect on the derived fluxes. Bright (and extended) foreground sources in close proximity to our high-redshift sample, however, are prone to imperfect modeling in this case (even with perfect PSFs) and leave residuals that in turn can significantly affect the photometry of faint sources nearby.

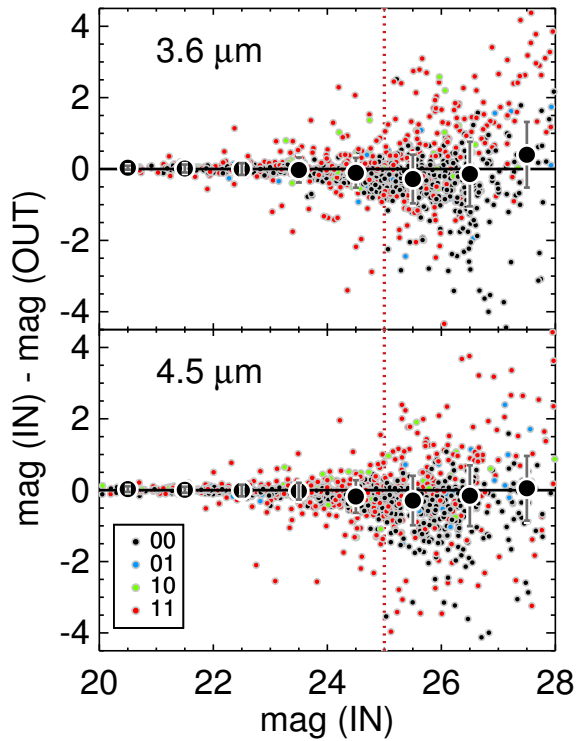

Fig. 3.— Comparison between the input and recovered IRAC 3.6 $\mu$m (upper) and 4.5 $\mu$m (lower) magnitude in our mock source simulations. Symbols are color-coded by blendedness in the two filters (shown in the inset in binary notation)—00: contamination-free; 01: contaminated in 4.5 $\mu$m; 10: contaminated in 3.6 $\mu$m; 11: contaminated in both 3.6 and 4.5 $\mu$m. Large black circles and error bars represent the median and standard deviation of the magnitude difference of sources in each magnitude bin with the magnitude difference between the input and the recovered less than 2 mag, demonstrating the reliability of our IRAC photometry down to the 50% completeness limit of 25 mag (red vertical dotted line; Ashby et al. 2015).

To account for these uncertainties, we visually inspected the IRAC science and residual images of all ~7000 sources in our sample to ensure that their IRAC photometry is reliable. Sources falling on strong residuals from nearby bright sources often have recovered S/Ns that are too high (even when we cannot visually identify counterparts in the IRAC images) or significantly negative, which indicates that their photometry is not reliable in general. This is confirmed by a mock source simulation in which we added mock sources at various positions around a bright source, finding that the recovered flux is highly biased (either overestimated or underestimated depending on the "yin" and "yang" of the residual on which the mock source was inserted). We therefore caution against blindly taking IRAC fluxes from a catalog and stress the importance of visual inspection of IRAC images and residuals of all objects, as contaminated sources can significantly impact studies on individual galaxies or with small number statistics (e.g., the high-mass end of the GSMF).

Contaminated sources are excluded from the sample in our subsequent analysis. This leaves ∼63%, 63%, 54%, 61%, and 57% of our $z = 4, 5, 6, 7, 8$ parent sample free from a possible contamination from nearby bright sources,[16] resulting in 2611, 1292, 364, 172, and 44 sources in our final $z = 4, 5, 6, 7, 8$ selection. Among the final sample, 1172/2611, 480/1292, 108/364, 41/172, and 6/44 (45%, 37%, 30%, 24%, and 14%) sources at $z = 4, 5, 6, 7, 8$, respectively, have $\gtrsim 2\sigma$ detections at 3.6 $\mu$m, and 613/2611, 211/1292, 43/364, 12/172, and 1/44 (23%, 16%, 12%, 7%, and 2%) show S/N > 5. A final multiwavelength catalog was constructed by combining the *HST* catalog and the IRAC T-PHOT catalog for this final sample.

## 3. STELLAR POPULATION MODELING

We derived stellar masses and rest-frame UV luminosities for our sample galaxies by fitting the observed spectral energy distribution (SED) from the $B$, $V$, $i$, $I_{850}$, $z$, $Y_{098}$, $Y$, $J$, $JH_{140}$, $H$, 3.6 $\mu$m, and 4.5 $\mu$m data to the Bruzual & Charlot (2003) SPS models. We refer the reader to Finkelstein et al. (2012, 2015) for a detailed explanation of our modeling process. Briefly, we modeled star formation histories (SFHs) as exponentially declining ($\tau = 1$ Myr–10 Gyr), constant ($\tau = 100$ Gyr), and rising ($\tau = -300$ Myr, $-1$ Gyr, $-10$ Gyr). Allowable ages spanned a lower age limit of 1 Myr to the age of the universe at the redshift of a galaxy, spaced semilogarithmically, and metallicity ranged from 0.02 to 1 $Z_\odot$. We assumed the Calzetti et al. (2000) attenuation curve with $E(B - V) = 0.0$–0.8 ($A_V = 0.0$–3.2), and intergalactic medium (IGM) attenuation was applied using the Madau (1995) prescription. All the models were normalized to a total mass of 1 $M_\odot$.

One of the major sources of uncertainty in stellar mass measurements of high-redshift galaxies derived from broadband imaging data via SED fitting is the contribution of nebular emission. Spectroscopically measuring the contribution of strong emission lines (e.g., H$\alpha$, [O III]$\lambda\lambda$4959, 5007) to the broadband fluxes for $z \gtrsim 4$ galaxies is not currently feasible because they redshift into the mid-IR. Many efforts, however, have been made taking an alternative approach of investigating IRAC colors of spectroscopically confirmed galaxies at the redshift range such that only one of the first two IRAC bands is expected to be contaminated by a strong emission line (e.g., Shim et al. 2011; Stark et al. 2013). Together with an inference from the direct measurements of nebular lines of galaxies at lower redshift (e.g., $z \sim 3$; Schenker et al. 2013a), several studies claim that the contribution of nebular emission to the broadband fluxes for galaxies at high redshift, and thus to the inferred stellar mass, can be significant (e.g., Schaerer & de Barros 2010).

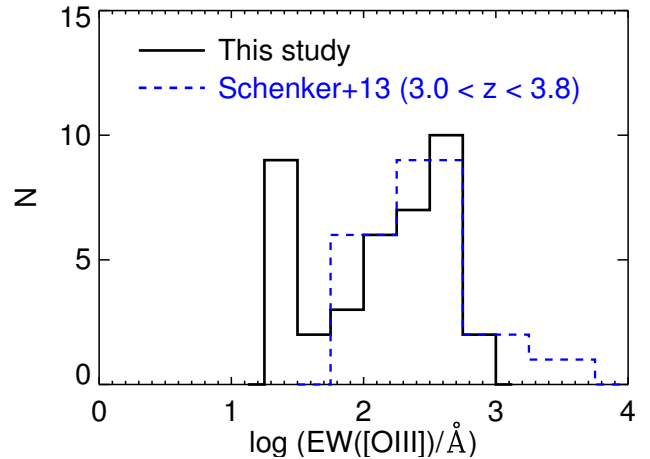

FIG. 4.— Comparison of the inferred rest-frame EW([O III]λ4959, 5007) distrbution derived from our SED fitting analysis for galaxies in our $z \sim 4$ sample with spectroscopic redshifts (black solid histogram) with the spectroscopic EW([O III]) distribution of 20 LBGs from Schenker et al. (2013a, blue dashed histogram; eight galaxies out of their 28 targets were not detected). The good agreement implies that our implementation of nebular emission lines in our stellar population models is a fair representation of reality.

While a more concrete answer on the nebular contribution will need to wait for the advent of the *James Webb Space Telescope* (*JWST*), we took into account the contribution of nebular emission in our SED modeling in a self-consistent way following the prescription of Salmon et al. (2015). In this prescription, the strengths of H and He recombination lines (including H$\beta$) are set by the number of non-escaping ionizing photons, and the strengths of metal lines relative to H$\beta$ are given by Inoue (2011). The nebular line strengths are thus a function of the population age and metallicity (which sets the number of ionizing photons), as well as the ionizing photon escape fraction. We added the nebular lines to the stellar continuum assuming an escape fraction of zero [17] and no extra dust attenuation for nebular emission (i.e., $E(B - V)_{\mathrm{stellar}} = E(B - V)_{\mathrm{nebular}}$). Figure 4 presents a comparison of the inferred equivalent width (EW) distribution of [O III]$\lambda\lambda$4959, 5007 from our SED modeling for spectroscopically confirmed galaxies at $3.5 < z < 4.5$ with the observed EW([O III]) histogram of 20 LBGs at similar redshifts ($3.0 < z < 3.8$) in Schenker et al. (2013a), showing a good agreement. As Figure 4 is based on spectroscopically confirmed galaxies, they are biased toward high-mass (median ± standard deviation of the logarithmic stellar mass $= 9.7\pm0.5$) and UV-bright ($-21.3\pm0.8$) galaxies. However, the two samples shown in Figure 4 have similar stellar mass distributions.

Our final stellar+nebular line stellar population mod-els are integrated through all of the filter bandpasses in our photometry catalog. The best-fit model for each source was found as the one that best represents the observed photometry via $\chi^2$ minimization. During this procedure, we accounted for uncertainties in the zero-point and aperture corrections by adding 5% of the flux in quadrature to the flux error in each band. For our fiducial stellar mass for each galaxy, we adopted the median mass obtained from a Bayesian likelihood analysis following Kauffmann et al. (2003a). We describe the procedure briefly here, but we refer the reader to Kauffmann et al. (2003a) and our previous work (Song et al. 2014) for more details of the analysis (see also Salmon et al. 2015; Tanaka 2015). In short, we used the $\chi^2$ array that samples the full model parameter space of our SPS models to compute the four-dimensional posterior pdf of free parameters (dust extinction, age, metallicy, and SFH) using the likelihood of each model, $\mathcal{L} \propto e^{-\chi^2/2}$. We assumed flat priors in parameter grids and $z = z_{\rm phot}$. The stellar mass for each grid point is taken to be the normalization factor between the observed SED and the model. Then, the 1-dimensional posterior pdf for stellar mass was obtained by marginalizing over all the parameters. The median and the central 68% confidence interval of stellar mass were computed from this marginalized pdf.

---

[16] These uncontaminated source fractions are consistent with the confusion-free fractions of the S-CANDELS images found by Ashby et al. (2015). While the symbols in Figure 3 are color-coded by the blendedness, which is determined based on the SExtractor segmentation map, in this section we visually inspected individual sources to exclude sources for which photometry is affected by residuals of nearby bright sources. That is, those excluded by our visual inspection are a subset of sources that are color-coded as being contaminated in Figure 3, and likely catastrophic outliers.

[17] Many studies at $z \sim 3$, the highest redshift where we can estimate the LyC escape fraction before the IGM becomes opaque to ionizing photons, have shown no evidence of high escape fraction, only placing an upper limit of $\lesssim 0.1$ ($1\sigma$; Siana et al. 2015).

The rest-frame absolute magnitude at 1500 Å, $M_{\rm UV}$, was obtained from the mean continuum flux density of the best-fit model in a $\Delta\lambda_{\rm rest} = 100$ Å band centered at rest-frame 1500 Å. Its uncertainty was derived from 100 Monte Carlo simulations in which the redshift uncertainty was accounted for by varying the redshift in our Monte Carlo simulations following the pdf, $P(z)$, obtained from our photometric redshift analysis (Finkelstein et al. 2015). Systematic biases and uncertainties of our SED fitting method were estimated via mock galaxy simulations and will be discussed in Section 4.3.

## 4. STELLAR MASS–REST-FRAME UV LUMINOSITY DISTRIBUTION

With the stellar mass ($M_*$) and the rest-frame absolute UV magnitude ($M_{\rm UV}$) of our galaxy sample measured from the previous section, we now explore the correlation between these properties in our sample to infer whether it is possible to derive a scaling relation.

### 4.1. Overall $M_*$–$M_{\rm UV}$ Distribution

Figure 5 presents the stellar mass versus rest-frame UV absolute magnitude distribution at each redshift. Overall, we find a strong trend between stellar mass and rest-frame absolute UV luminosity at all redshifts probed in this study. The scatter (standard deviation) in logarithmic stellar mass is about 0.4 dex (0.52, 0.42, 0.36, 0.40, and 0.30 dex at $z = 4$, 5, 6, 7, and 8, respectively, measured as the mean of standard deviation in logarithmic stellar mass in bins with more than five galaxies), and no noticeable correlation of the scatter is found with redshift or UV luminosity. The scatter at the bright end (measured at $-21.5 < M_{\rm UV} < -20.5$), where the effect of observational uncertainty should be minimal, is 0.43, 0.47, 0.36, 0.52, and 0.40 dex at $z = 4$, 5, 6, 7, and 8, respectively, similar to the quoted value above and the scatter at the faint end (at $-19.0 < M_{\rm UV} < -18.0$) of 0.51, 0.39, 0.39, and 0.41 dex at $z = 4$, 5, 6, and 7, respectively.

Often raised as a weakness in studies of rest-frame UV-selected galaxies is that such studies by construction miss dusty star-forming or quiescent galaxies. As our sample is selected mainly from rest-frame UV, the observed $M_*$–$M_{\rm UV}$ distribution shown in Figure 5 is also susceptible to this weakness. Interestingly, however, we do observe populations of UV-faint galaxies with high mass (given their UV luminosity) on the upper right part of the $M_*$–$M_{\rm UV}$ plane at $z = 4$ and 5. Although the lower limit of SFRs inferred from the UV luminosity and the Kennicutt (1998) conversion assuming no dust (upper x-axis in Figure 5) indicates that they are not completely quenched systems, their inferred (dust-uncorrected) SFRs are down to 2 orders of magnitude lower than those on the $M_*$–$M_{\rm UV}$ relation (to be derived in the following section). Outliers off the best-fit relation by more than 1 dex make up 6% (155/2624) of the total sample at $z = 4$. The fraction decreases as redshift increases to 3% (12/365) at $z = 6$. This increasing fraction of massive and UV-faint galaxy populations from $z = 6$ to $z = 4$ means either that we may be witnessing the formation of dusty star-forming or quiescent populations that are very rare at high redshift ($z \sim 6$–7) or that the duty cycle of those populations at high redshift is lower than that at low redshift (with a star-forming time scale much longer than 100 Myr) so that fewer such galaxies are observed with the current flux limit at higher redshift. Given the young age of the universe at high redshift, the latter requires a very early and fast growth of stellar mass for those galaxies that are completely quenched or highly dust extincted by $z \sim 7$. As we do not see such extremely UV-luminous populations at higher redshifts in the UV luminosity functions (e.g., Finkelstein et al. 2015; Bouwens et al. 2015), we regard the former as a more plausible scenario.

Interestingly, we do not find such populations in the opposite (low-mass) side at a given UV luminosity. The lack of bright and low-mass galaxies in the lower left region of Figure 5 is most clearly shown at $z = 4$. This is unlikely to be a selection effect or observational uncertainty, as had there been galaxies with $\log(M_*/M_\odot) \gtrsim 9$, they should have been detected in both WFC3/IR and IRAC; although we impose an S/N cut in both $J$ and $H$ bands in our sample selection and may thus be biased against the bluest galaxies, this only applies for UV bins fainter than those discussed here. It should not be an artifact of our SPS modeling, as the minimum mass-to-light ratio allowed in our SPS models is well below the mass-to-light ratio distribution of our sample. Lee et al. (2012), based on LBGs selected at $z = 4$–5 over the GOODS field, interpreted the absence of undermassive galaxies as evidence of smooth growth for UV-bright galaxies that has lasted at least a few hundred million years. If dust extinction is proportional to UV luminosity (e.g., Bouwens et al. 2014), this indicates a lack of UV-bright galaxies with very high specific SFRs (sSFR $=$ SFR$/M_*$). This lack of UV-bright and low-mass galaxies at all redshifts we probe provides a further support on our claim in the paragraph above of a growing population of dusty star-forming or quiescent galaxies seen between $z = 6$ and $z = 4$.

### 4.2. Stacking Analysis

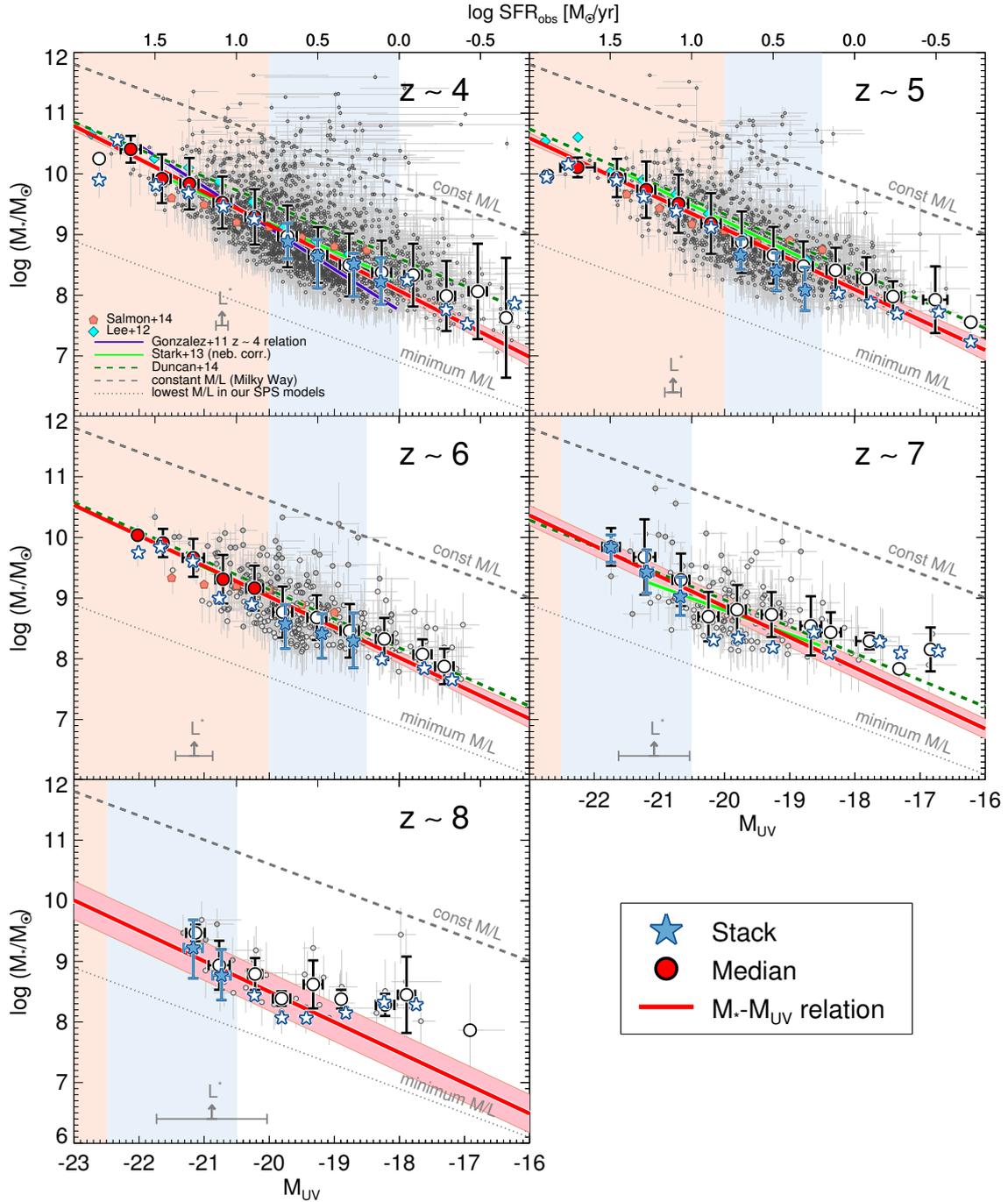

Fig. 5.— From *upper left* to *lower right*, stellar mass versus rest-frame UV absolute magnitude at 1500 Å ($M_{UV}$) at $z = 4$–8. For reference, SFR inferred from the UV luminosity and the Kennicutt (1998) conversion assuming no dust is shown in the upper x-axis. Small gray filled circles indicate objects with IRAC detection ($\gtrsim 2\sigma$ at 3.6 μm), while gray open circles are those with nondetections in IRAC ($< 2\sigma$ at 3.6 μm). Gray error bars represent the 68% confidence intervals in stellar mass and rest-frame absolute UV magnitude. Large red filled circles are the median stellar masses in each rest-frame absolute UV magnitude bin of 0.5 mag, while large open circles indicate bins containing a single galaxy. Black error bars are standard deviations in stellar mass in each UV luminosity bin. Blue stars indicate median stellar masses in each rest-frame absolute UV magnitude bin from our median-flux stacking analysis in Section 4.2, with error bars denoting the $1\sigma$ uncertainty, including both photometric error and sample variance. We derived the best-fit relation (red solid line) by fitting data points that combine red filled circles with blue filled stars in a redshift-dependent UV magnitude range specified in Section 4.3 (indicated as the light-red and light-blue filled regions). The $1\sigma$ uncertainty of the best-fit $M_\star$–$M_{UV}$ relation is denoted as the light-red shaded region. The gray arrows and horizontal error bars at the bottom show the characteristic UV magnitude, $L^*$, of the UV luminosity function (Finkelstein et al. 2015) at each redshift. Gray dotted lines indicate minimum mass-to-light ratio allowed in our SPS models. We find that the best-fit relation has a nonevolving slope at $z = 4$–6, which is marginally steeper than a constant mass-to-light ratio (gray dashed line; normalized to the mass-to-light ratio of the Milky Way), and shows a weak evolution in the normalization. The inferred stellar mass from the best-fit $M_\star$–$M_{UV}$ relation for galaxies with $\bar{M}_{UV} = -21$ ($\sim L^*$) is $\log(M_\star/M_\odot) = 9.70, 9.59, 9.53, 9.36,$ and 9.00 at $z = 4, 5, 6, 7,$ and 8, respectively. The typical mass-to-light ratio of galaxies at $z = 4$–8 at the rest-frame UV absolute magnitude of the Milky Way ($M_{UV} = -20.5$) is lower by a factor of ∼30 (130) at $z = 4$ (8) than that of the Milky Way.

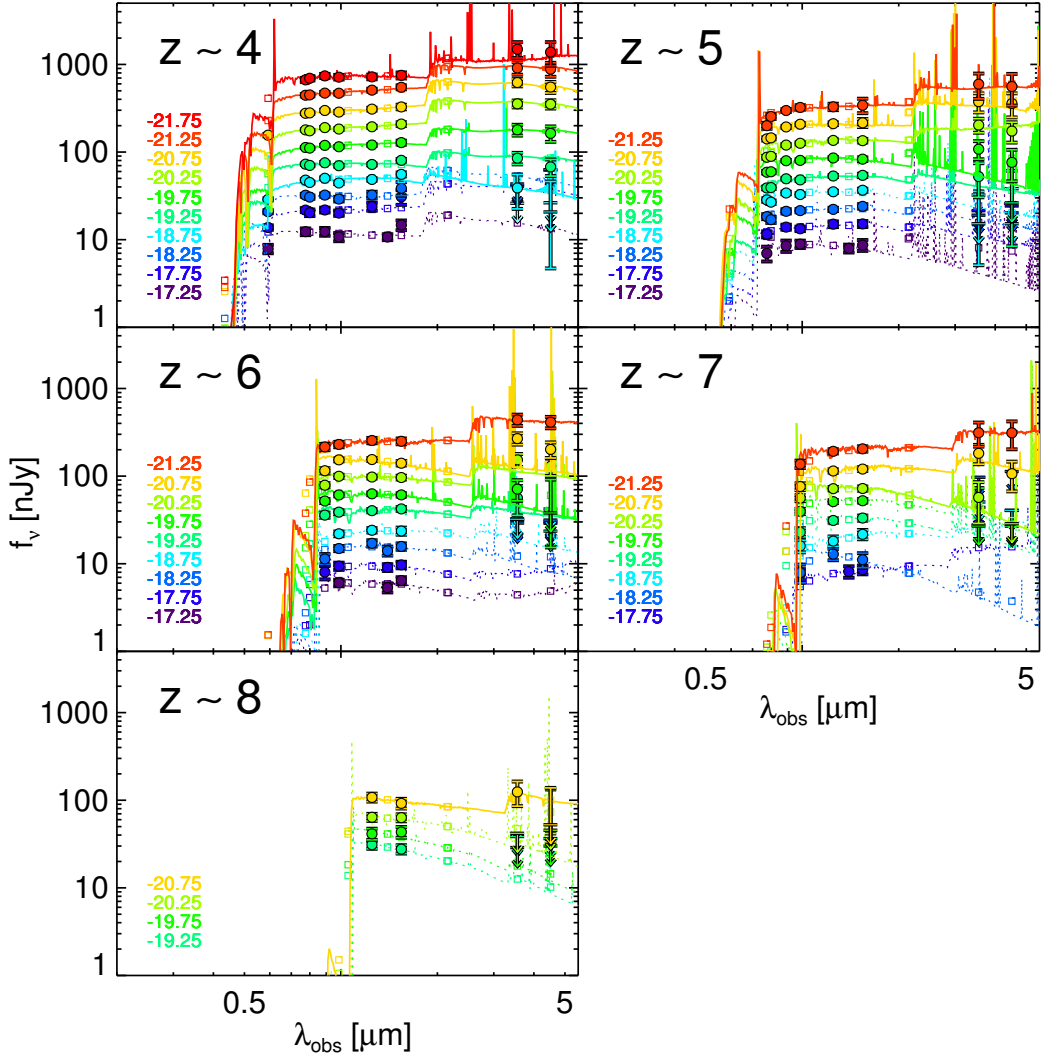

FIG. 6.— Median flux-stacked SEDs at $z = 4$–$8$ in bins of rest-frame absolute UV magnitude with $\Delta M_{UV} = 0.5$ mag, for bins with $M_{UV} < -17$ and more than 10 (for $z = 4$–5) or 5 (for $z = 6$–8) galaxies (corresponding to blue stars in Figure 5). The stacked SEDs are denoted by filled circles and downward-pointing arrows (indicating $2\sigma$ upper limits for bands with S/N < 2), with the rest-frame absolute UV magnitudes given by the inset text. The solid lines and open squares indicate the best-fit SPS models and model bandpass-averaged fluxes, respectively. The best-fit SPS models for stacked SEDs with S/N < 2 in 3.6 $\mu$m are shown as dotted lines, indicating that the inferred mass for stacked SEDs with S/N < 2 in 3.6 $\mu$m is uncertain.

Despite our deep IRAC data, individual galaxies in our sample, especially those in faint UV luminosity bins (which likely have low masses), often suffer from low S/N in the IRAC mosaics. This can make the reliability of an $M_*$–$M_{UV}$ relation derived based on individual galaxies questionable. We therefore performed a stacking analysis to increase the S/N and to examine the typical stellar mass in each rest-frame absolute UV magnitude bin. We built median flux-stacked SEDs, comprising a total of 12 bands ($B$, $V$, $i$, $I_{814}$, $z$, $Y_{098}$, $Y$, $J$, $JH_{140}$, $H$, 3.6 $\mu$m, 4.5 $\mu$m), for galaxies in each UV magnitude bin of 0.5 mag in the full sample. Uncertainties on the stacked SEDs were assigned as the quadrature sum of the photometric uncertainty and the uncertainty due to sample variance (heterogeneity of the SEDs of galaxies) estimated via bootstrapping on galaxies in each UV magnitude bin. The latter dominates the error budget at $z = 4$, contributing on average 80% to the total uncertainty, but the contribution of sample variance decreases with redshift, down to 45% at $z = 8$. This is a combined effect of (i) decreasing outliers in the $M_*$–$M_{UV}$ plane (i.e., decreasing fraction of dusty star-forming or quiescent galaxy population) with redshift, as seen in Figure 5, and (ii) increasing photometric uncertainty with redshift, as the galaxies are fainter compared to the photometric depth.

Stacked SEDs were analyzed through our SED fitting procedures described in Section 3 with the redshift of model templates fixed to the median photometric redshift of galaxies in each stack. Bands not common to all galaxies (e.g., $JH_{140}$ covering only the HUDF) were included in the SED fitting only when more than half of the stacked galaxies have measurements. Our choice is a trade-off between making the most of the available information and minimizing the chance of biasing our results by including a subset not having the characteristics of the parent sample. As the median flux-stacked SEDs shortward of the Ly$\alpha$ line may still have some fluxes depending on the redshift distribution of the stacked galaxies and thus may not represent an SED of a galaxy at the median redshift, we excluded bands shortward of the Ly$\alpha$ line.

The best-fit SPS models and stacked SEDs in each redshift bin are shown in Figure 6. Overall, the shapes of the SEDs are nearly similar but show a weak trend with UV luminosity such that the *typical* UV-bright galaxies have slightly redder rest UV-to-optical (observed NIR-to-MIR) color than UV-faint galaxies at a given redshift, being in qualitative agreement with other previous studies (e.g., González et al. 2012). This trend indicates that on average UV-faint galaxies have (mildly) lower mass-to-light ratios than UV-bright galaxies, suggesting that an $M_*$–$M_{\rm UV}$ relation with a constant mass-to-light ratio would not provide a good description of our data.

The results from our SED fitting analysis on the median flux-stacked SEDs are included in the $M_*$–$M_{\rm UV}$ plots of Figure 5. Comparing the stacked points to the median of individual galaxies shows that they are largely consistent with each other at the bright end ($M_{\rm UV} \lesssim -(20$–$21)$). But for fainter UV luminosities, the stacked points are generally lower than the medians. This may reflect the fact that the stellar masses of individual IRAC-undetected galaxies are on average biased toward higher masses, while they are relatively robust for IRAC-detected galaxies.

### 4.3. Mock Simulation with Semi-Analytic Models

As noted in the previous section, stellar mass estimation involves various sources of uncertainty, which impact the derived GSMFs. In our methodology of convolving the $M_*$–$M_{\rm UV}$ distribution with the completeness-corrected UV luminosity functions to derive GSMFs, the reliability of our to-be-derived GSMF and its low-mass-end slope is tied to our ability to recover the *intrinsic* slope, normalization, and scatter of the $M_*$–$M_{\rm UV}$ distribution.

To explore the systematics and uncertainties in our observed $M_*$–$M_{\rm UV}$ distribution introduced by the photometric uncertainty of our data and our stellar mass estimation procedure, and to examine whether we can recover the *intrinsic* $M_*$–$M_{\rm UV}$ distribution (and the subsequent GSMF), we simulated $M_*$–$M_{\rm UV}$ planes using mock galaxies drawn from semianalytic models (SAMs). We used synthetic galaxy photometry from the SAMs of Somerville et al. (2016, in preparation). These SAMs are implemented on halos extracted from the *Bolshoi* dark matter simulation (Klypin et al. 2011), which has a dark matter mass resolution of $1.35 \times 10^8\ h^{-1} M_\odot$ and a force resolution of $1\ h^{-1}$ kpc. Galaxies hosted in a halo more massive than $10^{10}\ M_\odot$ are included in the mock catalog. This corresponds to a typical stellar mass of a few times $10^7\ M_\odot$ at $z = 4$–8 (e.g., Behroozi et al. 2013), similar to the minimum stellar mass in our real sample. The most notable characteristic of these SAMs is that their light cones are specifically designed to provide realizations of the five CANDELS fields (with albeit a factor of 6–9 larger areas than the actual CANDELS *HST* coverage), aiming to help with interpretation of observational data. Moreover, using synthetic galaxy photometry from the SAMs has an advantage over using SPS models in that mock galaxies have more realistic SEDs based on more complicated SFHs and metal enrichment histories, thus representing real galaxy populations more closely. We refer the reader to Somerville et al. (2012) and Lu et al. (2014) for full details of their mock galaxy models. Here we specifically use the mock light cone of the CANDELS GOODS-S field for our simulation.

We generated mock galaxy samples by populating the $M_*$–$M_{\rm UV}$ plane at each redshift with objects from the SAM catalog similar to our real sample in both sample size and rest-frame UV absolute magnitude distribution, but with various input slopes ranging from $-0.3$ to $-0.8$. We assumed a lognormal distribution around a linear $\log M_*$–$M_{\rm UV}$ relation with a dispersion of $\sim0.3$ dex, motivated by the functional form of the observed star-forming main sequence at lower redshifts (Speagle et al. 2014 and references therein). Although the SAMs have an inherent M/L relation, this does not affect our results as we randomly draw galaxies from the SAM to fill in our simulated plane based on the M/L slope in a given simulation. We then added noise in each band based on the flux uncertainty of our real sample at a given magnitude and perturbed the simulated photometry assuming a Gaussian error distribution. The stellar masses and rest-frame UV absolute magnitudes of these mock galaxies were calculated in the same manner as in the real data. The above precedures describe a single mock realization of one intrinsic $M_*$–$M_{\rm UV}$ distribution. For each realization of a given intrinsic $M_*$–$M_{\rm UV}$ distribution, we measured the recovered slope, normalization, and scatter of the $M_*$–$M_{\rm UV}$ distribution. We repeated the above procedures 50 times for each input slope and redshift, from which we constructed pdfs of the recovered slopes and normalizations.

In addition to allowing us to explore the best way to minimize the bias and uncertainty in recovering the intrinsic $M_*$–$M_{\rm UV}$ distribution, these simulations also allow us to test the validity of our stellar mass measurements. First, we find that if we use the classical maximum likelihood estimator (i.e., the best-fit model) as a fiducial stellar mass, the uncertainty in the inferred stellar mass for individual galaxies is considerable, differing up to 1 dex for galaxies with $\log(M_*/M_\odot) \sim 9$ at $z = 4$. As a result, the scatter of the recovered $M_*$–$M_{\rm UV}$ distribution increases significantly compared to the original one (a 0.2 dex increase at $M_{\rm UV} \sim -20$ and larger for fainter galaxies; see also Salmon et al. 2015). This large spread in stellar mass in the recovered distribution makes it hard to recover the intrinsic relation at $z \gtrsim 6$ where the photometric uncertainty is large and the sample size is small. The median mass from our Bayesian likelihood analysis described in Section 3 performs much better in the sense that it does not significantly increase the scatter of the recovered $M_*$–$M_{\rm UV}$ plane even in faint UV luminosity bins; the scatter of the recovered $M_*$–$M_{\rm UV}$ distribution compared to that of the input distribution shows an increase of only 0.05–0.10 dex at $M_{\rm UV} \sim -20$, and the scatter remains nearly constant in fainter UV luminosity bins. However, there exists a noticeable bias in the recovered stellar mass for galaxies fainter (and lower in mass) than a certain UV magnitude threshold at each redshift (hereafter referred to as $M_{\rm thresh1}(z)$). This is because, for low-S/N data, the stellar mass is determined by the assumed priors (e.g., flat priors in parameter grids

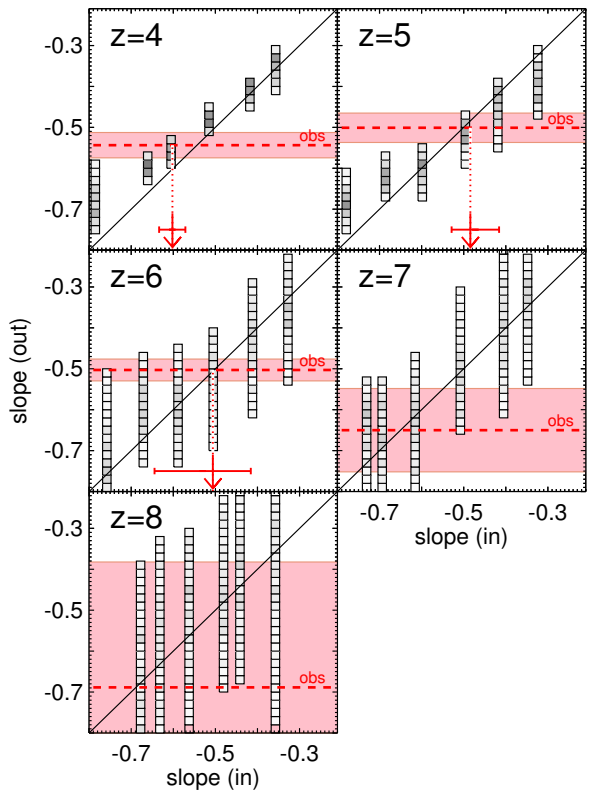

Fig. 7.— Probability density function of the recovered $M_*$–$M_{UV}$ slope as a function of the intrinsic slope at $z = 4$–$8$ in our SAM mock galaxy simulations. Darker gray colors indicate a higher probability. For reference, a 1:1 line is shown as the black solid line. The best-fit slope of the $M_*$–$M_{UV}$ relation and its $1\sigma$ uncertainty obtained from our *real* sample (using our optimized fitting method) at each redshift are denoted as the red dashed horizontal line and shaded region, respectively. At $z = 4$–$6$, we can recover the intrinsic slope within $\pm 0.07$. At $z = 7$–$8$, the uncertainties become much larger; thus, as discussed in Section 4.4, we fix the slope (which does not significantly evolve from $z = 4$ to $6$) at $z = 7$ and $8$ to the $z = 6$ value.

in the case of our SPS modeling) and the data have little constraining power.

Although the recovered $M_*$–$M_{UV}$ distribution above $M_{\mathrm{thresh1}}(z)$ is reliable in both bias and scatter, the small dynamic range above this limit at high redshift still makes the uncertainty of the recovered $M_*$–$M_{UV}$ relation large. We thus utilize a stacking analysis (described in Section 4.2) to increase the S/N and dynamic range in UV luminosity and to reliably derive typical properties of sources (e.g., stellar mass). In the simulation, we find that via stacking, we can achieve this 1.5–2.0 mag further in UV magnitude, down to $M_{UV} \sim -(18.0$–$18.5)$ at $z = 4$–$6$ (hereafter $M_{\mathrm{thresh2}}(z)$). $M_{\mathrm{thresh2}}(z)$ corresponds roughly to the UV luminosity of the stacked SEDs with S/N $\sim 1$ at $3.6\ \mu$m, below which the inferred mass remains uncertain just as one might expect.

In short, our simulation enables us to assess the bias and uncertainty of the observed $M_*$–$M_{UV}$ distribution, which has so far been generally overlooked in the literature. We find that the observed $M_*$–$M_{UV}$ distribution and the inferred relation at high redshift are very sensitive to the choice of stellar mass estimator and the UV magnitude range even when using the same data set and can be dominated by systematics if not tested thoroughly. From this simulation, we derive the redshift-dependent UV magnitude thresholds, $M_{\mathrm{thresh1}}(z)$ and $M_{\mathrm{thresh2}}(z)$, above which we can rely on the median mass and scatter of individual galaxies in each UV magnitude bin and the median mass of stacks, respectively. Specifically, we find $M_{\mathrm{thresh1}}(z)$ to be $M_{UV} = -20.0, -20.0, -20.0, -22.5, -22.5$ and $M_{\mathrm{thresh2}}(z)$ to be $M_{UV} = -18.0, -18.5, -18.5, -20.5, -20.5$ at $z = 4, 5, 6, 7, 8$.

Based on our findings, we derive the best-fit $M_*$–$M_{UV}$ relation by combining the median mass of galaxies in each UV magnitude bin at $M_{UV} < M_{\mathrm{thresh1}}(z)$ and the median mass of stacked points at $M_{\mathrm{thresh1}}(z) < M_{UV} < M_{\mathrm{thresh2}}(z)$, neglecting galaxies fainter than $M_{\mathrm{thresh2}}(z)$ when fitting this relation. To explore the validity of this optimized method of combining individual bright galaxies with stacked faint galaxies, we show in Figure 7 the distribution of the recovered slope at $z = 4$–$8$ for various input slopes with our optimized method, demonstrating that we can recover the intrinsic relation fairly well even at $z = 6$. From this simulation, we derive the pdf of intrinsic slope (slope$_{\mathrm{in}}$) of the $M_*$–$M_{UV}$ relation given the observed slope (slope$_{\mathrm{out}}$), based on the results from each of the 50 simulations at each input slope. We find that, for the given slope observed from our real sample, the central 68% range of the intrinsic slope to be $-0.57 <$ slope$_{\mathrm{in}} < -0.63$, $-0.42 <$ slope$_{\mathrm{in}} < -0.53$, and $-0.42 <$ slope$_{\mathrm{in}} < -0.64$ at $z = 4$, $5$, and $6$, respectively. This is comparable with the $1\sigma$ confidence level of the observed slope, indicating that our $M_*$–$M_{UV}$ relations and their uncertainties at $z = 4$–$6$ measured from our real sample are relatively reliable, when we use this optimized method.

It is encouraging that we find no bias at $z = 7$–$8$, but the broad pdf of the recovered slope for a given input slope at $z = 7$–$8$ indicates that it is hard to constrain the intrinsic slope with the currently available data. Therefore, we perform another test to see if our current data can constrain the normalization of the intrinsic $M_*$–$M_{UV}$ relation when we apply an additional prior of a known intrinsic slope. Figure 8 shows the pdf of the recovered normalization as a function of intrinsic normalization at

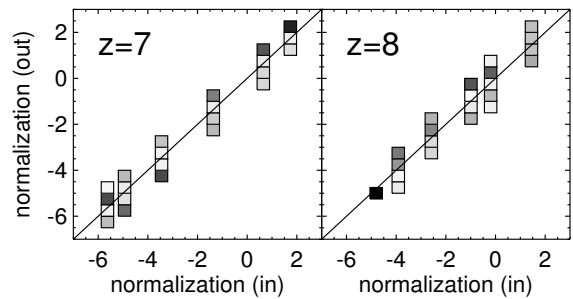

Fig. 8.— Probability density function of the recovered $M_*$–$M_{UV}$ normalization as a function of intrinsic normalization at $z = 7$ and $z = 8$ in our mock galaxy simulations, when the slope is fixed to the intrinsic slope. When the slope is fixed, the normalization can be recovered at high confidence.

| $z$ | Normalization ($M_\odot$) | Slope | $\log M_{*(M_{\rm UV}=-21)}$ ($M_\odot$) |
|---|---|---|---|
| 4 | $-1.70 \pm 0.65$ | $-0.54 \pm 0.03$ | $9.70 \pm 0.02$ |
| 5 | $-0.90 \pm 0.74$ | $-0.50 \pm 0.04$ | $9.59 \pm 0.03$ |
| 6 | $-1.04 \pm 0.57$ | $-0.50 \pm 0.03$ | $9.53 \pm 0.02$ |
| 7 | $-1.20 \pm 0.16$ | $-0.50 \pm$ — | $9.36 \pm 0.16$ |
|   | $(-4.23 \pm 2.12)$ | $(-0.65 \pm 0.10)$ | $(9.45 \pm 0.07)$ |
| 8 | $-1.56 \pm 0.32$ | $-0.50 \pm$ — | $9.00 \pm 0.32$ |
|   | $(-5.47 \pm 6.19)$ | $(-0.69 \pm 0.31)$ | $(9.00 \pm 0.31)$ |

Note. — At $z \sim 7$–8, numbers in parentheses are the best-fit parameters when we do not fix the slope to the best-fit slope at $z \sim 6$. The quoted errors represent the $1\sigma$ uncertainties. The normalization is defined as the logarithmic stellar mass at $M_{\rm UV} = 0$ ($\log M_{*(M_{\rm UV}=0)}$).

$z = 7$–8 when the slope is fixed to the intrinsic slope, illustrating that the normalization can be recovered accurately within $\pm 0.5$.

The best-fit relation for our real sample in Section 4.4 is derived with the optimized method of combining individual bright galaxies with stacks of fainter galaxies. When we need to use the full $M_*$–$M_{\rm UV}$ distribution or a scatter for our GSMF derivation, we use those inferred at $M_{\rm UV} < M_{\rm thresh1}(z)$.

### 4.4. $M_*$–$M_{\rm UV}$ Relation

Using our optimized method vetted in Section 4.3, we derive the best-fit linear $\log_{10}(M_*/M_\odot)$–$M_{\rm UV}$ relation at each redshift. Specifically, we use the median mass of individual galaxies for bins with $M_{\rm UV} < M_{\rm thresh1}(z)$ and the median mass of median flux-stacked SEDs for bins with $M_{\rm UV} < M_{\rm thresh2}(z)$. Our best-fit mass-to-light relation, listed in Table 1, has a slope of $-(0.50$–$0.69)$, which is slightly steeper than a constant mass-to-light ratio of a slope of $-0.40$. As the slope between $z = 4$ and $z = 6$ is nearly constant ($\sim -0.5$) but the uncertainty at $z = 7$–8 is large, we fix the slope at $z = 7$ and $z = 8$ to be the same as the best-fit slope at $z = 6$ while leaving the normalization as a free parameter.[18] Our mock galaxy simulation in Section 4.3 indicates that our current data allow us an accurate recovery of the intrinsic normalization at $z = 7$–8 with a prior of a known input slope (Figure 8). We find a (very) weak evolution in the normalization between $z = 4$ and $z = 7$ (a decreasing normalization with increasing redshift). The normalization evolves from $\log(M_*/M_\odot)_{(M_{\rm UV}=-21)} = 9.70$ (at $z = 4$) to 9.36 (at $z = 7$) at only $2\sigma$ significance. Interestingly, the normalization appears to decrease more rapidly from $z = 7$ to $z = 8$ than at lower redshifts, although the small sample at $z = 8$ prevents drawing any firm conclusions. We discuss this further in Section 6.

Although the $M_*$–$M_{\rm UV}$ distribution of our flux-limited sample has nonzero scatter, the derived $M_*$–$M_{\rm UV}$ relation is not subject to Malmquist bias (i.e., missing faint galaxies at a given stellar mass) as we estimate the $M_*$–$M_{\rm UV}$ relation in bins of luminosity and not stellar mass. Therefore, the derived relation should be robust against the Malmquist bias, which could artificially result in a steeper slope than the intrinsic one by losing the faint envelope of galaxy distribution for a given stellar mass.[19]

There are discrepancies of 0.3–0.7 dex between different studies in the measured median mass at a given UV magnitude even at $z \sim 4$, which are larger at fainter UV bins (González et al. 2011; Lee et al. 2012; Stark et al. 2013; Duncan et al. 2014; Salmon et al. 2015). This may reflect a number of systematic uncertainties associated with sample selection and stellar mass estimation. As these uncertainties make a direct comparison between different studies difficult, it highlights the importance of a comprehensive and independent analysis to verify systematics.

Overall, we found the best-fit slope shallower than that of González et al. (2011) but similar to those of Stark et al. (2013) and Duncan et al. (2014), with an exception at $z = 4$. First, the slope of our best-fit relation of $-0.54(\pm 0.03)$ at $z = 4$ is significantly shallower than that of González et al. of $-0.68(\pm 0.08)$, which is based on an order-of-magnitude-smaller sample. As the relation of González et al. is derived with no nebular correction, it is not surprising that their stellar masses for UV-bright galaxies are higher than ours. However, their stellar masses for galaxies in faint UV bins are lower than ours by $\sim 0.2$ dex at $M_{\rm UV} \sim -18$, resulting in a steeper slope than ours. Meanwhile, the $M_*$–$M_{\rm UV}$ relation of Stark et al. (2013) shows a good agreement with ours at all redshifts with only a slightly shallower slope than ours. Duncan et al. (2014) in general find higher stellar masses for galaxies than we do in faint UV bins, resulting in a higher normalization and a shallower slope than ours, with the biggest difference in normalization ($\Delta \log(M_*/M_\odot)_{M_{\rm UV}=-21} = 0.25$) and slope ($\Delta = 0.09$; $2.5\sigma$ significance) being observed at $z = 4$ (see Section 5.4 for more discussion).

## 5. THE GALAXY STELLAR MASS FUNCTION

### 5.1. Derivation of the GSMFs

We now convolve our $M_*$–$M_{\rm UV}$ distribution with the observed rest-frame UV luminosity function to derive GSMFs. The luminosity functions we utilize in this study are from Finkelstein et al. (2015), which included the full CANDELS GOODS fields, the HUDF and two parallel fields, and the Abell 2744 and MACS J0416.1-2403 parallel fields from the Hubble Frontier Fields data set. These luminosity functions are already corrected for incompleteness and selection effects using the detection probability kernels derived from fake source simulations, as discussed by Finkelstein et al. (2015).

Figure 9 presents GSMFs constructed using four different methods:

1. "Raw Bootstrapped GSMF":

We constructed the "observed" GSMF by combining the UV luminosity function with the observed $M_*$–$M_{\rm UV}$

---

[18] Albeit slightly shallower ($-(0.44$–$0.48)$), the slope predicted from the SAMs of Somerville et al. (2016, in preparation) is almost constant over the redshift interval as well.

[19] As discussed in Section 4.1, objects with high mass and low UV flux that are observed or that we may be missing are believed to be the subdominant population for all UV luminosity bins and thus would not change the derived relation (see also Salmon et al. 2015 for a similar argument based on the tight scatter of the star-forming main sequence). When probing in bins of stellar mass, however, those massive outliers seen at $z = 4$–5 with $\log(M_*/M_\odot) > 10$ are off from the linear correlation derived from in bins of UV luminosity in this section.

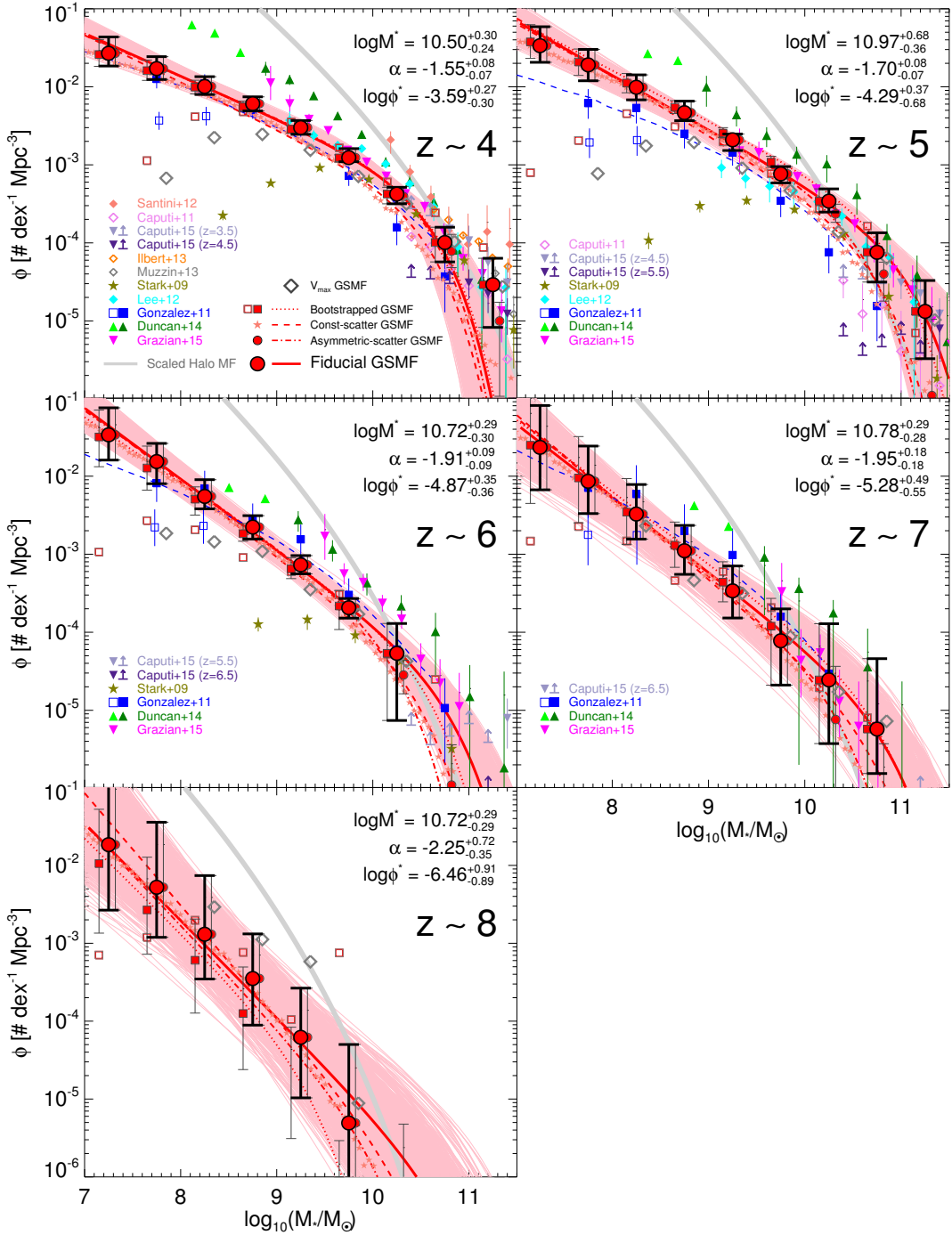

FIG. 9.— Our fiducial galaxy stellar mass functions, from a $J + H$-band selected sample of galaxies at $z = 4$–8 (red filled circles). The open squares, filled squares, small stars, and small red filled circles indicate raw bootstrapped, incompleteness-corrected bootstrapped, constant-scatter, and asymmetric-scatter GSMFs, respectively. The red dotted, red dashed, red dash-dot and red solid lines represent the Schechter fit for the last three and for our fiducial GSMFs, respectively. The uncertainty on our fiducial GSMF includes contributions from both the UV luminosity function uncertainties and the uncertainty in the $M_\star$–$M_{UV}$ relation. The light-red shaded regions denote 1,000 Schechter fits for our fiducial GSMF randomly chosen within the $1\sigma$ 3-dimensional contour of the Schechter parameters determined from our MCMC analysis. The blue points (open squares, filled squares, dashed line) correspond to previous estimates (raw bootstrapped, incompleteness-corrected bootstrapped, constant-scatter GSMF) of González et al. (2011) from WFC3/IR data of the ERS (for $z = 4$–6) and from WRC3/IR data of the ERS, HUDF09, and NICMOS data over the GOODS fields (for $z = 7$). Also overplotted are recent estimates for GSMFs from the literature—from $K$s-band selected sample at $z \sim 4$ (Ilbert et al. 2013; Muzzin et al. 2013), from 4.5$\mu$m selected sample (Caputi et al. 2011, 2015), and from rest-frame UV-selected samples (Stark et al. 2009; Santini et al. 2012; Lee et al. 2012; Duncan et al. 2014; Grazian et al. 2015). All points and lines are converted to a Salpeter IMF. The thick grey lines show dark matter halo mass functions *scaled to a baryon conversion efficiency of 20%* (i.e., 20% of halo mass times the cosmic baryon fraction of $\Omega_b/\Omega_m$). Our GSMFs are characterized by a steeper low-mass-end slope of $-1.55$ ($-2.25$) at $z = 4$ ($z = 8$) compared to that of González et al. of $-1.43$ ($-1.55$) at $z = 4$ ($z = 7$).

distribution. We first drew $10^5$ galaxies from the best-fit Schechter function of the UV luminosity function in the range of $-30 < M_{UV} < -13$. Then, we assigned stellar mass for each galaxy to be the stellar mass of the randomly chosen galaxy with similar rest-frame UV absolute magnitude from the observed $M_*$–$M_{UV}$ distribution. This method accounts for outliers, which may be a non-negligible fraction of galaxies at $z = 4$.

The non-negligible spread in stellar mass at a fixed rest-frame UV absolute magnitude observed in Figure 5 can result in GSMFs with an underestimated low-mass-end slope if we do *not* account for the "unobserved" population of galaxies below the detection threshold of the survey. This is because galaxies at a given UV luminosity have a range of mass-to-light ratios; thus, galaxies below the detection limit can still have stellar masses high enough to contribute to the number density in the stellar mass bins of our interest (affecting the last few points in the GSMFs depending on the amount of spread). Therefore, we need to correct for the incompleteness by assuming an $M_*$–$M_{UV}$ distribution in UV luminosity bins below the current sensitivity limit. We use three different incompleteness correction schemes.

2. "Incompleteness-corrected Bootstrapped GSMF":

A reasonable assumption on the $M_*$–$M_{UV}$ distribution in the unobserved faint UV luminosity bins is that the observed distribution of individual galaxies in bright UV bins above a threshold represents the intrinsic distribution and can be extended to fainter bins. In Section 4.3, we found the redshift-dependent threshold $M_{\mathrm{thresh1}}(z)$ at each redshift above which the intrinsic distribution can be recovered well without any noticeable bias or increase in scatter. We extend the observed $M_*$–$M_{UV}$ distribution at $M_{UV} < M_{\mathrm{thresh1}}(z)$ to fainter UV luminosities down to $M_{UV} = -13$, keeping the distribution centered around the best-fit $M_*$–$M_{UV}$ relation. The rest of the procedures are the same as those for the "raw bootstrapped GSMF".

3. "Constant-scatter GSMF":

Instead of using individual points in the $M_*$–$M_{UV}$ distribution, we assumed an idealized lognormal distribution around the best-fit $M_*$–$M_{UV}$ relation with a constant scatter, inferred from bright UV luminosity bins that have a statistical number of galaxies and high completeness. We used the mean scatter estimated from the two faintest bins at $M_{UV} < M_{\mathrm{thresh1}}(z)$ with more than 10 galaxies, which is $\sim$0.4–0.5 dex. The UV luminosity function was then convolved with this lognormal distribution to derive the "constant-scatter GSMF".

4. "Asymmetric-scatter GSMF":

The three approaches above are basically the same as those used by González et al. (2011). However, the mass distribution of our sample in a given rest-frame UV absolute magnitude bin is not symmetric with respect to the best-fit relation. Rather, the lower side of the best-fit relation has in general a smaller scatter. As already noted in Section 4.1, the lack of galaxies with high UV luminosity and low mass contributes in part to the asymmetric scatter, and the results in Section 4.3 in addition indi-cate that it could be an intrinsic property and not just an observational bias. Therefore, we assume a lognormal distribution with an asymmetric scatter with respect to the best-fit $M_*$–$M_{UV}$ relation (a different sigma above and below the mean) inferred from the two faintest bins at $M_{UV} < M_{\mathrm{thresh1}}(z)$ and extend it to fainter UV luminosity bins.[20] The resultant GSMF constructed via this method is referred to as the "asymmetric-scatter GSMF".

While the asymmetric-scatter GSMFs were devised to probe the incompleteness-corrected low-mass-end slope, they might not account properly for the fraction of outliers seen at $z = 4$ and $z = 5$ that can impact the high-mass end of the GSMF. Thus, we combine the bootstrapped GSMF for high masses ($\log(M_*/M_\odot) > 10$) and the asymmetric-scatter GSMFs for low masses ($\log(M_*/M_\odot) > 10$), and consider this our fiducial GSMF. We found that the GSMFs computed using the $1/V_{\max}$ method (Schmidt 1968, gray diamonds in Figure 9) are in excellent agreement with our fiducial GSMFs at the high-mass end. Figure 9 shows our fiducial GSMFs at $z = 4$–8 (listed in Table 2) as well as those derived from the four methods described above. The incompleteness-corrected, constant-scatter, and asymmetric-scatter methods yield consistent results (except at $z = 8$ where the GSMF is less robust due to the small sample size), and their Schechter fits (see Section 5.2) are nearly indistinguishable within the uncertainties on the low-mass end.

### 5.2. Schechter Fit and Uncertainties

Before we parameterize our GSMF with a Schechter function, we first need to derive the uncertainties for our GSMF data points. Randomly perturbing points of the GSMFs is not a proper way to estimate the uncertainties of the GSMFs because they are correlated. Thus, we derive the 68% confidence interval of our fiducial GSMFs as follows. First, we randomly drew 1000 samples from a Markov Chain Monte Carlo (MCMC) chain of the UV luminosity function Schechter parameters derived by Finkelstein et al. (2015) within the 3-dimensional $1\sigma$ contour of $(L^*, \alpha_L, \phi_L^*)$. Then, for each luminosity function generated from the Schechter parameters, we assigned an $M_*$–$M_{UV}$ relation with (slope, normalization) values randomly chosen within the 2-dimensional $1\sigma$ contour of the best-fit parameters of the $M_*$–$M_{UV}$ relation.[21] The new $M_*$–$M_{UV}$ distribution was then combined with the new luminosity function to generate a GSMF in the same way that our GSMF was constructed. This generates 1000 GSMFs, of which the minimum and maximum represent the $1\sigma$ upper and lower limits of the GSMFs, respectively.

We parameterize our GSMFs with a Schechter (1976) function,

$$\phi(M_*)dM_* = (\phi^*/M^*) \times (M_*/M^*)^\alpha \exp[-(M_*/M^*)]dM_* \tag{1}$$

---

[20] For $z = 7$ and $z = 8$ where the scatter is not robustly measured due to the small sample size, we assume that the $M_*$–$M_{UV}$ distribution follows that at $z = 6$.

[21] For $z = 7$–8 where we fix the slope of the $M_*$–$M_{UV}$ relation to the $z = 6$ value, the uncertainty in the normalization is taken into account.

which is characterized by a power law with a low-mass-end slope of $\alpha$, an exponential cutoff at stellar masses larger than a characteristic mass, $M^*$, and a normalization $\phi^*$. The best fit, uncertainties, and posterior distributions of the Schechter parameters for our fiducial asymmetric-scatter GSMFs were derived by running an MCMC algorithm that samples the 3-dimensional parameter space of the Schechter parameters. To ensure full coverage of the parameter space and assess convergence, we ran five parallel chains composed of $10^5$ steps each. The starting position of the chain was determined by a coarse grid search that minimized the $\chi^2$ statistic between the model and the data. The first 10% of steps were disregarded in the burn-in phase before running each chain to reduce the dependence of the posterior distribution on the initial position. The proposal distribution of each parameter was assigned as a normal (lognormal) distribution for $\alpha$ ($M^*$ and $\phi^*$) with the width tuned to generate an acceptance rate of $\sim$ 23–30%.[22] As a prior, we limited the sampling parameter space to be $\alpha > -10$, $8 < \log(M^*/M_\odot) < 13$, and $\log(\phi^*/\mathrm{Mpc}^{-3}) > -8$. For $6 \leq z \leq 8$, where the constraints on $M^*$ are weak (see the large error bars on the open gray circles in Figure 10), we took a lognormal prior on $M^*$ with mean $\log(M_*/M_\odot) = 10.75$ and standard deviation of 0.3 dex, following the posterior distribution of $M^*$ at $z = 4$ and $z = 5$. Comparing the likelihood of the current step, which is defined as $\mathcal{L} \propto e^{-\chi^2/2}$, with the proposed set of parameters determines whether the proposal is accepted. We employ the Metropolis–Hastings algorithm for the acceptance criteria. After running all chains, convergence was assessed by examining trace plots of parameters, as well as using the Rubin–Gelman $\hat{R}$ diagnostic (Gelman & Rubin 1992) for each marginalized posterior distribution. The diagnostic value $\hat{R} \sim 1$ suggests convergence, and we confirmed that for all redshifts and parameters the diagnostic has a value $1.00 < \hat{R} < 1.01$.

From the resulting joint posterior distribution, we extracted the marginal posterior distribution of each parameters. The median and the central 68% of the marginal posterior distributions provide our fiducial Schechter parameters and an estimate of the 68% confidence interval on each parameter (shown as the error bars in Figure 10). In short, the uncertainties on the GSMFs and Schechter parameters include (i) the uncertainty of the best-fit $M_*$–$M_{\mathrm{UV}}$ relation and (ii) the uncertainty of the Schechter parameters in the UV luminosity function, the latter of which includes Poisson errors. Other sources of random errors on the derived GSMFs, including cosmic variance, are discussed in Section 7.2.

Our best-fit Schechter parameters are listed in Table 3 and plotted in Figure 10 as a function of redshift. Our data support a decreasing (steepening) of the low-mass-end slope ($\alpha \sim -1.55^{+0.08}_{-0.07}$, $-1.70^{+0.08}_{-0.07}$, $-1.91^{+0.09}_{-0.09}$, $-1.95^{+0.18}_{-0.18}$, $-2.25^{+0.72}_{-0.35}$ at $z = 4, 5, 6, 7, 8$) with increasing redshift, asymptoting to the faint-end slope of the UV luminosity function ($\alpha_L \sim -1.56^{+0.06}_{-0.05}$, $-1.67^{+0.05}_{-0.06}$,

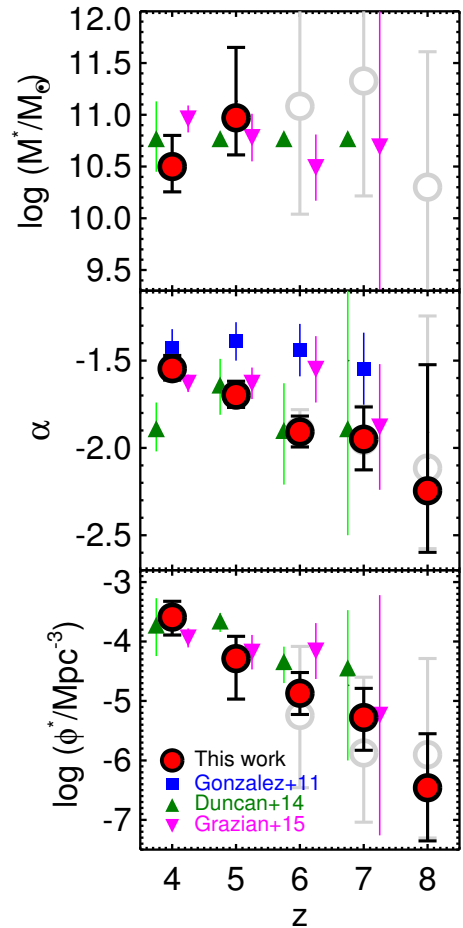

Fig. 10.— Redshift evolution of the best-fit Schechter parameters for our fiducial GSMFs. The low-mass-end slope, $\alpha$, evolves toward a steeper value with increasing redshift, asymptoting to the faint-end slope of the UV luminosity function. Conversely, the open gray circles denoting the best-fit Schechter parameters with a flat prior on $M^*$ show that we observe no significant evolution in the characteristic mass, $M^*$, though our observations do not allow us to place reasonable constraints on $M^*$ at $z \geq 6$.

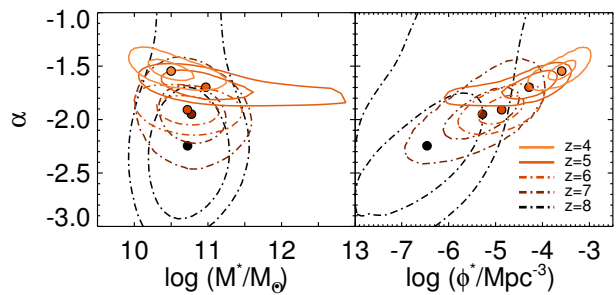

Fig. 11.— Confidence contours of the best-fit Schechter parameters for our fiducial GSMFs at $z = 4$–$8$ at the 68% and 95% levels, showing that we can place reasonable constraints on $\alpha$. The best-fit Schechter parameters at $z = 6$–$8$ were derived with a lognormal prior on $M^*$, and their contours are denoted as dot-dashed lines. The best-fit values are shown as filled circles.

$-2.02^{+0.10}_{-0.10}$, $-2.03^{+0.21}_{-0.20}$, $-2.36^{+0.54}_{-0.40}$ at $z = 4, 5, 6, 7, 8$), and possibly a decrease in $\phi^*$ with increasing redshift as well. Conversely, our GSMFs favor no evolution in $M^*$ with redshift, although we cannot rule out the possiblity of evolution due to the large uncertainties, in particular at $z \gtrsim 6$ (see the gray open circles in Figure 10, which are derived with a flat prior on $M^*$). The large error bars on $M^*$ are because constraints on the bright end of the luminosity function (and thus the massive end of the GSMF) remain weak, as even our wide-area data reach only 1–2 mag brighter than the characteristic magnitude of the UV luminosity function, $L^*$ (marked as arrows at the bottom of Figure 5), and the UV magnitude bins brighter than $L^*$ are populated by only a few galaxies. The lack of robust constraints on the massive end of the GSMF leads to a well-known degeneracy between $M^*$, $\alpha$, and $\phi^*$, as shown in Figure 11 with confidence contours on the Schechter parameters.

### 5.3. *Eddington Bias*

Several recent studies (Ilbert et al. 2013; Caputi et al. 2015; Grazian et al. 2015) investigated the effect of the Eddington bias (Eddington 1913) on the low-mass-end slope of the GSMF, finding that the increase in the stellar mass uncertainties for individual galaxies at low masses results in an artificial steepening of the low-mass-end slope (Grazian et al. 2015; Caputi et al. 2015). In our methodology, the "effective" uncertainty in stellar mass is not stellar mass dependent, as we do not use individual stellar masses to construct the GSMF on the low-mass end but rather convolve the UV LF with stellar mass–UV luminosity distribution, where we use the distribution around ~$L^*$ ($-21 < M_{UV} < -20$) to extend to fainter luminosities. Taking a similar approach to other studies, we investigated the effect of the Eddington bias by parameterizing the observed GSMF as a convolution of an "intrinsic" Schechter function with a lognormal function with its width to be the mean stellar mass uncertainty of galaxies with $-21 < M_{UV} < -20$. We implemented a grid search for the "intrinsic Schechter parameter to correct for the Eddington bias, finding that while we see a similar trend to what Grazian et al. (2015) found (the "intrinsic" low-mass-end slope is similar [at $z = 4$–5] or slightly flatter [at $z \gtrsim 6$], driven by the change in $M^*$, which is uncertain), the change in the number density on the low-mass end is negligible and the trend of steepening low-mass-end slope with increasing redshift persists after correcting for the Eddington bias.

### 5.4. *Comparison with Previous Work*

Figure 9 compares our GSMFs with recent estimates from the literature determined by rest-frame UV-selected galaxies (either using photometric redshift or color-color selection; Stark et al. 2009; González et al. 2011; Lee et al. 2012; Santini et al. 2012; Duncan et al. 2014; Grazian et al. 2015)[23] or rest-frame optical-selected galaxies (Caputi et al. 2011; Ilbert et al. 2013; Muzzin et al. 2013; Caputi et al. 2015). All GSMFs are converted to a Salpeter IMF when necessary. Overall, the comparison of our GSMFs with previous estimates demonstrates that there

---

[23] For the GSMFs of Stark et al. (2009), we apply correction factors for nebular emission of ×1.1, 1.3, and 1.6 at $z = 4, 5, 6$ that are inferred in Stark et al. (2013).

---

exists a considerable discrepancy in the high- and/or low-mass ends and in the normalization of the GSMFs at all redshifts. However, before discussing the discrepancy in detail, we point out that the error bars of the GSMFs compared in Figure 9 include only random uncertainties. For example, those of Grazian et al. (2015) include Poisson errors and errors (photometric scatter and photometric redshifts) derived from their Monte Carlo simulations, while those of Duncan et al. (2014) include Poisson errors and photometric redshift uncertainty. Those of González et al. (2011) include the uncertainties of the UV luminosity function of Bouwens et al. (2007, 2011) and the scatter of the $M_*$–$M_{UV}$ relation. None of the plotted points include cosmic variance or other sources of systematic uncertainties.

A major source of discrepancy in the GSMFs between different studies may be attributed to the systematic uncertainties associated with stellar mass estimation, as already noted by many studies (e.g., Marchesini et al. 2009; Mobasher et al. 2015). Each study adopts different assumptions on the SFH, dust law, metallicity, nebular emission, etc., in their SPS modeling, of which effects on the derived stellar mass can be significant. Moreover, many parameters are degenerate, thus making assessment of the systematic effects induced by these different assumptions on the disagreement of the observed GSMFs hard to achieve. While investigating the systematic effects caused by the different sets of assumptions adopted in previous studies is beyond the scope of our study, we stress that we have focused on deriving the *intrinsic* GSMFs by exploring the systematic effects inherent in our analysis, minimizing them via our SAM+mock galaxy simulations (see Section 4.3).

With these above caveats in mind, we now discuss the discrepancy highlighted by the direct comparison between the GSMFs in Figure 9. First, we observe a disagreement in the normalization of the GSMFs between ours and the estimates from the literature at all redshifts. At $z = 4$ and $z = 5$, although our normalization is in good agreement with Lee et al. (2012) and Grazian et al. (2015), a prominent discrepancy is observed with González et al. (2011), Santini et al. (2012), and Duncan et al. (2014). The GSMFs of González et al. (2011) are found to lie systematically below ours and others in the literature at $z = 4$ and $z = 5$, and the GSMFs of Santini et al. (2012) at $z \sim 4$ and Duncan et al. (2014) at all redshifts are found to lie above. At higher redshifts of $z = 6$–7, the normalization of our GSMFs is lower with respect to those of Duncan et al. (2014) and Grazian et al. (2015, at $z = 6$) but shows a good agreement with Stark et al. (2009) (at the massive end; they do not correct for incompleteness), González et al. (2011), and Grazian et al. (2015, at $z = 7$). Although the overall difference in the normalization of the GSMFs between different studies remains similar at $z = 7$, the larger error bars at $z = 7$ render any differences at that high redshift insignificant.

Examining the different mass regimes, we notice an interesting disagreement at $z = 4$, where wide-area ground-based surveys (Ilbert et al. 2013; Muzzin et al. 2013), which are potentially more sensitive to more rare, massive galaxies, may be more robust. Specifically, in the most massive bin of our study at $\log(M_*/M_\odot) = 11.25$, both Muzzin et al. (2013) and Ilbert et al. (2013) found a number density ~ 0.1–0.3 dex higher than we find

TABLE 2
Galaxy stellar mass function at $z = 4$–$8$

| log $M_*$ ($M_\odot$) | log $\phi$ (dex$^{-1}$ Mpc$^{-3}$) | | | | |
|---|---|---|---|---|---|
| | $z = 4$ | $z = 5$ | $z = 6$ | $z = 7$ | $z = 8$ |
| 7.25 | $-1.57^{+0.21}_{-0.16}$ | $-1.47^{+0.24}_{-0.21}$ | $-1.47^{+0.35}_{-0.32}$ | $-1.63^{+0.54}_{-0.54}$ | $-1.73^{+1.01}_{-0.84}$ |
| 7.75 | $-1.77^{+0.15}_{-0.14}$ | $-1.72^{+0.20}_{-0.20}$ | $-1.81^{+0.23}_{-0.28}$ | $-2.07^{+0.45}_{-0.41}$ | $-2.28^{+0.84}_{-0.64}$ |
| 8.25 | $-2.00^{+0.13}_{-0.10}$ | $-2.01^{+0.16}_{-0.16}$ | $-2.26^{+0.21}_{-0.16}$ | $-2.49^{+0.38}_{-0.32}$ | $-2.88^{+0.75}_{-0.57}$ |
| 8.75 | $-2.22^{+0.09}_{-0.09}$ | $-2.33^{+0.15}_{-0.10}$ | $-2.65^{+0.15}_{-0.15}$ | $-2.96^{+0.32}_{-0.30}$ | $-3.45^{+0.57}_{-0.60}$ |
| 9.25 | $-2.52^{+0.09}_{-0.09}$ | $-2.68^{+0.07}_{-0.14}$ | $-3.14^{+0.12}_{-0.11}$ | $-3.47^{+0.32}_{-0.35}$ | $-4.21^{+0.63}_{-0.78}$ |
| 9.75 | $-2.91^{+0.12}_{-0.05}$ | $-3.12^{+0.09}_{-0.11}$ | $-3.69^{+0.12}_{-0.13}$ | $-4.11^{+0.41}_{-0.57}$ | $-5.31^{+1.01}_{-1.64}$ |
| 10.25 | $-3.37^{+0.09}_{-0.12}$ | $-3.47^{+0.16}_{-0.14}$ | $-4.27^{+0.38}_{-0.86}$ | $-4.61^{+0.72}_{-0.82}$ | — |
| 10.75 | $-4.00^{+0.20}_{-0.25}$ | $-4.12^{+0.25}_{-0.38}$ | — | $-5.24^{+0.90}_{-0.57}$ | — |
| 11.25 | $-4.54^{+0.34}_{-0.55}$ | $-4.88^{+0.40}_{-0.61}$ | — | — | — |

Note. — The quoted 1$\sigma$ errors include the uncertainties of the UV luminosity function and the $M_*$–$M_{UV}$ relation.

TABLE 3
Best-fit Schechter function parameters of our fiducial GSMFs

| $z$ | log $M^*$ ($M_\odot$) | $\alpha$ | $\phi^*$ ($10^{-5}$ Mpc$^{-3}$) |
|---|---|---|---|
| 4 | $10.50^{+0.30}_{-0.24}$ | $-1.55^{+0.08}_{-0.07}$ | $25.68^{+21.75}_{-12.80}$ |
| 5 | $10.97^{+0.68}_{-0.36}$ | $-1.70^{+0.08}_{-0.07}$ | $5.16^{+7.05}_{-4.08}$ |
| 6 | $10.72^{+0.29}_{-0.30}$ | $-1.91^{+0.09}_{-0.09}$ | $1.35^{+1.16}_{-0.75}$ |
| | $(11.08^{+1.27}_{-1.04})$ | $(-1.90^{+0.12}_{-0.54})$ | $(0.57^{+7.70}_{-0.54})$ |
| 7 | $10.78^{+0.29}_{-0.28}$ | $-1.95^{+0.18}_{-0.18}$ | $0.53^{+1.10}_{-0.38}$ |
| | $(11.33^{+1.05}_{-1.11})$ | $(-1.96^{+0.20}_{-0.17})$ | $(0.13^{+2.37}_{-2.12})$ |
| 8 | $10.72^{+0.29}_{-0.29}$ | $-2.25^{+0.72}_{-0.35}$ | $0.035^{+0.246}_{-0.030}$ |
| | $(10.30^{+1.31}_{-1.08})$ | $(-2.12^{+0.87}_{-0.46})$ | $(0.125^{+5.057}_{-0.125})$ |

Note. — The quoted best-fit values and 1$\sigma$ errors of the Schechter parameters represent the median and the central 68% confidence interval of the marginal posterior distribution of each parameter obtained from our MCMC analysis. At $6 \leq z \leq 8$, we show in parentheses the results derived with a flat prior on $M^*$.

(and $\sim$0.2–0.3 dex higher than Duncan et al. (2014) and Grazian et al. (2015)). This discrepancy may be attributed in part to the fact that the median redshift of these ground-based surveys is $z = 3.5$, lower than ours, as well as the other studies shown ($z \simeq 4$). Moreover, the former is derived from a shallow but wide ($\sim$1.6 deg$^2$ down to $K_s \sim 24$) $K_s$-band (rest-frame optical) selected catalog. Thus, they should be more complete and less susceptible to cosmic variance and Poisson noise at the high-mass end than other works that are based on small-field, rest-frame UV-selected catalogs. Meanwhile, Caputi et al. (2015), who investigated the high-mass end of the GSMFs by adding the contribution of $K_s$-band faint but 4.5$\mu$m bright ([4.5] < 23) galaxies to the previous determinations of the $K_s$-band-selected GSMFs by Caputi et al. (2011) and Ilbert et al. (2013), present consistent results with ours, which may be attributed to the lower normalization of the GSMFs of Caputi et al. (2011) with respect to others.

At the high-mass end, GSMFs based on rest-frame UV-selected galaxies (Stark et al. 2009; González et al. 2011; Lee et al. 2012; Duncan et al. 2014; Grazian et al. 2015) agree reasonably well with each other when the cosmic variance is accounted for (see Section 7.2.2). However, the high-mass end at $z = 4$ from Santini et al. (2012) still shows a mild tension with ours due to their overall higher normalization with respect to others.

Turning to the low-mass end, while our survey volume is smaller than those of the ground-based surveys of Ilbert et al. (2013) and Muzzin et al. (2013), the deep data set in this study enables us to reach lower in mass than these surveys can (log($M_*/M_\odot$) $\gtrsim$ 10), allowing more robust constraints on the low-mass-end slope. Starting at $z = 4$, our results at the low-mass end are consistent with most previous rest-frame UV-selected studies (González et al. 2011; Lee et al. 2012; Duncan et al. 2014; Grazian et al. 2015), with the exception of Duncan et al. (2014).

Because Grazian et al. (2015) restrict their analysis to higher masses, the only points for comparison at log($M_*/M_\odot$) $\lesssim$ 9 are those from González et al. (2011) and Duncan et al. (2014). Interestingly, the largest disagreement at the low-mass end is found at the lowest redshift of $z = 4$, where Duncan et al. (2014) find significantly higher number densities ($\sim$0.5 dex at log($M_*/M_\odot$) $\sim$ 9) and a steeper low-mass-end slope ($\alpha \sim -1.9$) with respect to the others. This may result from differences in the measured $M_*$–$M_{UV}$ relation, as well as in the faint-end slope of the UV luminosity function between our study and that of Duncan et al. (2014). At $z = 4$, Duncan et al. (2014) found a shallower $M_*$–$M_{UV}$ slope than what we find here. One difference in methods is that rather than using our hybrid approach of using individual high-mass galaxies and stacks of lower-mass galaxies, Duncan et al. (2014) fit their $M_*$–$M_{UV}$ distribution over a wide stellar mass range down to log($M_*/M_\odot$) $\sim$ 8. As shown in their simulations (see their Fig. 5), stellar masses for galaxies with log($M_*/M_\odot$) < 9 are biased toward higher masses (a similar result to what we find here), leading to their derivation of a shallower $M_*$–$M_{UV}$ slope. A shallower slope of the $M_*$–$M_{UV}$ relation translates into a steeper low-mass-end slope and a higher normalization of the GSMF: for a given number density of galaxies in bins of UV luminosity ($\phi_L$), the number density of galaxies in bins of stellar mass ($\phi_M$) is given by $\phi_M \propto \phi_L (dL/dM)$, and is thus higher for a shallower $M_*$–$M_{\rm UV}$ slope. We do note, however, that Duncan et al. (2014) did not use this $M_*$–$M_{\rm UV}$ relation to measure their GSMF; they used a $1/V_{\rm max}$ method. However, the biases inherent in measuring masses from individual poorly detected galaxies may still be responsible for their steeper low-mass-end slope at $z = 4$. Moreover, their faint-end slope of the UV luminosity function at $z \sim 4$ is steeper (by $\sim$0.2) than that from Finkelstein et al. (2015), on which our GSMF is based.

Unsurprisingly, differences are thus found in the evolution of the low-mass-end slope with redshift. The Schechter fit for our GSMF indicates steeper low-mass-end slopes. While González et al. found a tentative steepening in the low-mass-end slope with increasing redshift, the steepening is mild, from $-1.43$ at $z = 4$ to $-1.55$ at $z = 7$. This is a combined effect of a steeper faint-end slope of the updated UV luminosity function by Finkelstein et al. (2015) used in our analysis and a shallower $M_*$–$M_{\rm UV}$ relation found in this study (Section 4.2). Grazian et al. (2015) and Duncan et al. (2014) both find a nearly constant low-mass-end slope of $\alpha = -1.6$ and $\alpha = -1.9$, respectively, and no evidence of the steepening that we observe.

At lower redshifts of $z < 4$, the consensus is that the characteristic mass does not change but the normalization evolves (e.g., Marchesini et al. 2009), although the evolution of the low-mass-end slope remains controversial as some find no evolution (Marchesini et al. 2009) while others find a steepening low-mass-end slope with increasing redshift (for a single Schechter function fit; Kajisawa et al. 2009; Santini et al. 2012; Ilbert et al. 2013; Tomczak et al. 2014). In this study, at $z \gtrsim 4$, the observed evolution of the GSMF shows a low-mass-end slope that steepens with redshift. Additionally, our results tentatively confirm a roughly constant $M^*$, with a decreasing $\phi^*$ with increasing redshift, qualitatively similar to results at lower redshift (e.g., Ilbert et al. 2013), though we acknowledge that our relatively small volume prevents robust constraints on $M^*$.

## 6. STELLAR MASS DENSITY

To measure the stellar mass density at $z = 4$–8, we integrated the best-fit Schechter function at each redshift from $8 < \log(M_*/M_\odot) < 13$, an often adopted interval for stellar mass density estimates at high redshift in the literature. Table 4 lists our estimates of the stellar mass density along with their $1\sigma$ uncertainties, calculated as the minimum and maximum stellar mass densities allowed by the 3-dimensional $1\sigma$ contour of the Schechter parameters obtained in Section 5.2. Figure 12 presents the evolution of the stellar mass density at $z = 4$–8, alongside values compiled from the literature (converted to a Salpeter IMF when necessary). Most data points from the literature are taken from the compilation by Madau & Dickinson (2014), with the exception of González et al. (2011), which Madau & Dickinson (2014) corrected for nebular emission. We instead show the uncorrected (original) points together with Stark et al. (2013), which are the González et al. values corrected for nebular emission. We also show the recently published works of Duncan et al. (2014) and Grazian et al. (2015). The error bars from most of the studies include only ran-dom errors.

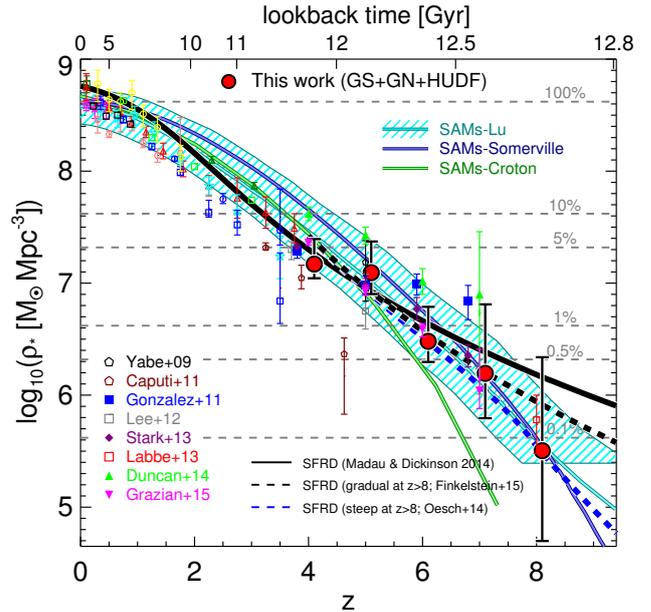

FIG. 12.— Evolution of the stellar mass density. The stellar mass densities were obtained by integrating the best-fit Schechter functions for our fiducial GSMFs between $M_* = 10^8 \ M_\odot$ and $M_* = 10^{13} \ M_\odot$ (red circles). The error bars indicate the minimum and maximum values of stellar mass density allowed by the $1\sigma$ contour of the Schechter parameters. Small symbols are the compilation of stellar mass densities from the literature by Madau & Dickinson (2014) (using their colors and symbols) along with recent estimates from Stark et al. (2013), Duncan et al. (2014), and Grazian et al. (2015), listed in the legend. All points and lines are converted to a Salpeter IMF. The solid black curve marks parameterization of the time integral of SFRD from Madau & Dickinson (2014) after gas recycling ($R = 0.27$) is accounted for, representing the prediction for the stellar mass density. The stellar mass densities predicted for the two scenarios suggested by Oesch et al. (2014) for the SFRD evolution at $z > 8$ are denoted as the black and blue dashed lines (see Section 6 for more details). Our stellar mass densities show a remarkable agreement with estimates of the stellar mass density from the SFRD. Other colored solid curves are stellar mass densities predicted from the three SAMs introduced in Section 7.1.1, and the cyan hatched region denotes the 95% posterior range of the Lu et al. (2014) model. For reference, we denote fractions of the local stellar mass density measurement (Baldry et al. 2012) as horizontal dotted lines.

Our estimates of the stellar mass density at $z = 4$–5 are in broad agreement with previous measurements within the uncertainty, with the exception of Caputi et al. (2011) at $z = 5$ and Duncan et al. (2014) at $z = 4$. Duncan et al. found a $\sim$0.5 dex higher stellar mass density at $z = 4$, a deviation at $1.9\sigma$, mainly due to their steeper low-mass-end slope compared to ours. The stellar mass density at $z \sim 5$ from Caputi et al. shows a value lower by about 0.8 dex than our estimates (a deviation at $4.2\sigma$), which is surprising, given that their median redshift is slightly lower ($z \sim 4.6$). However, the sample selection of Caputi et al. is very different from ours, such that their sample is selected in the IRAC 4.5 $\mu$m band, which, while more complete for very red galaxies, may underestimate the incompleteness to star-forming galaxies.

At higher redshifts of $z = 6$–7, our measurements are $\sim$0.5–0.7 dex lower than those of González et al. (2011) and Duncan et al. (2014) but show an excellent consis-

| $z$ | $\log \rho_*$ |
|---|---|
| | $(M_\odot \, \mathrm{Mpc}^{-3})$ |
| 4 | $7.17 \, ^{+0.22}_{-0.13}$ |
| 5 | $7.09 \, ^{+0.28}_{-0.19}$ |
| 6 | $6.48 \, ^{+0.31}_{-0.18}$ |
| | $(6.53 \, ^{+0.55}_{-0.26})$ |
| 7 | $6.19 \, ^{+0.62}_{-0.40}$ |
| | $(6.24 \, ^{+0.88}_{-0.45})$ |
| 8 | $5.50 \, ^{+0.83}_{-0.81}$ |
| | $(5.51 \, ^{+1.64}_{-1.80})$ |

NOTE. — Stellar mass estimates by integrating the best-fit Schechter functions for our GSMFs over $8 < \log(M_*/M_\odot) < 13$. The quoted $1\sigma$ error bars represent the minimum and maximum values of stellar mass density allowed by the 3-dimensional $1\sigma$ contour of the Schechter parameters. The values in parentheses at $6 \le z \le 8$ are the results from the best-fit Schechter function derived with a flat prior on $M^*$, displaying negligible difference from our fiducial stellar mass estimates.

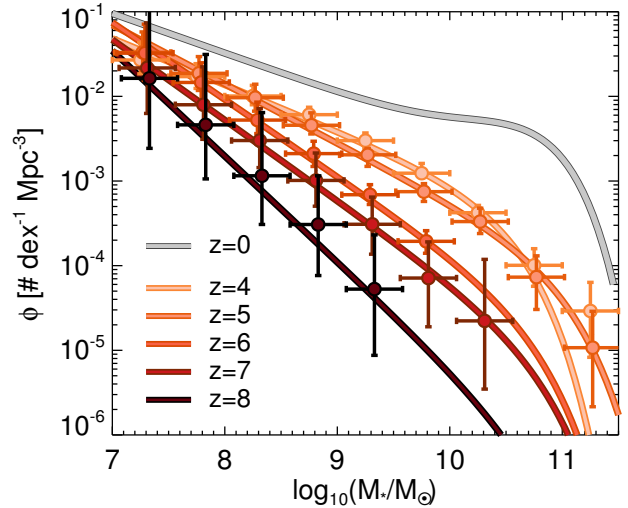

FIG. 13.— Redshift evolution of our fiducial GSMFs at $z = 4$–$8$. For reference, the gray thick line denoting the $z \sim 0$ GSMF (Baldry et al. 2012) is shown.

tency with the recent estimate of Grazian et al. (2015). The agreement in the stellar mass density in spite of the difference in the normalization of the GSMFs between our study and Grazian et al. (2015) at $z = 6$ (their higher normalization compared to ours) is a consequence of their shallower low-mass-end slope by 0.4, which compensates the difference in the stellar mass density due to their higher normalization when integrating the GSMFs. As the GSMFs are less well constrained at these high redshifts, our stellar mass density is only in mild tension with other studies at $z = 6$ ($< 2\sigma$). At $z = 7$, although the estimates of the stellar mass density from different studies differ up to 0.7 dex, the increased error bars mean that these differences are not currently statistically significant.

As the stellar mass is to first order the time integral of past star formation activity, a comparison of the stellar mass density with the time integral of the SFRD should yield similar values if both estimates are accurate. At high redshifts, however, both quantities have large uncertainties. The limiting factor in determining an accurate SFRD is a determination of dust attenuation, for which the observed UV luminosity density is corrected, without a direct observation of the dust-obscured star formation for most cases. Likewise, the uncertainties involved in the determination of the stellar mass density (the systematics in stellar mass estimates, the uncertainties on the abundance of low-mass galaxies, etc.) can impact the stellar mass density and potentially result in a mismatch between the integral of the SFRD and the stellar mass density.

Figure 12 compares the time integral of the SFRDs with the stellar mass density derived in this work. In Figure 12, the black solid line indicates the parameterization by Madau & Dickinson (2014) of the time integral of the SFRD.[24] The black dashed line shows the stellar mass density inferred by the SFRD parameterization of Finkelstein et al. (2015, $\log \mathrm{SFRD} \propto (1+z)^{-4.3}$), which used updated values for the SFRD at $z = 4$–$8$. Finally, the blue dashed line is from Oesch et al. (2014), who suggested from a dearth of $z > 8$ galaxy candidates that the SFRD appears to decline more rapidly at $z > 8$ ($\log \mathrm{SFRD} \propto (1+z)^{-10.9}$) than predicted from the evolutionary trend in the SFRD at lower redshifts of $4 < z < 8$, though given the large uncertainty in the SFRD estimates at $z > 8$, this claim currently remains controversial.

While at low redshift the expected stellar mass density from the SFRD systematically exceeds the observed stellar mass density by $\sim$0.3 dex (see Conroy 2013 for summary and discussion on recent improvements), at $4 < z < 7$ we do not observe such a trend. In particular, with the updated SFRD measurements from Finkelstein et al. (2015), the expected stellar mass density from the SFRD is in good agreement with our stellar mass densities within the uncertainties, which is a somewhat remarkable result given the potential systematic uncertainties in the measurements of both the SFRD and the stellar mass density. In a scenario in which galaxies undergo episodic star formation with a timescale longer than 100 Myr (the timescale traced by UV), they would have a UV-"dark" phase, and thus the stellar mass density would be lower than the time integral of the SFRD when both measurements are based on the UV-"bright" sample. Therefore, the agreement between the two estimates implies that on average the duty cycle of star formation in relatively massive star-forming galaxies is high and episodic accretion is not the dominant mode of star formation. This is also hinted at in Section 4.1 and is in agreement with other studies (e.g., Papovich et al.

_______
[24] All the time integral of the SFRDs presented in this paper account for a gas recycling fraction of $R = 0.27$ for a Salpeter IMF.

2011). Nonetheless, we cannot rule out short-term fluctuations (<100 Myr) in the SFH, which would still give an agreement between the two quantities if measured from the rest-frame UV-selected sample.

At $z = 8$, there is an intriguing steep drop in the stellar mass density, which results in it being consistent with the steep dropoff in the SFRD inferred by Oesch et al. (2014), although still consistent with a smooth extrapolation from the SFRD evolution from Finkelstein et al. (2015) and Madau & Dickinson (2014). However, the constraints on the stellar mass density at $z = 8$ are weak, due to the large uncertainties in the $M_*$–$M_{UV}$ relation, which is based on two stacked points consisting of only 11 galaxies. Thus, a larger sample of $z = 8$ galaxies, combined with deeper IRAC imaging, is necessary to robustly measure the $z = 8$ stellar mass density.

Our estimates imply that the stellar mass density has increased by a factor of $10^{+30}_{-8}$ from $z = 7$ to $z = 4$, and 0.4%, 0.7%, 3.0%, and 3.5% of the present-day stellar mass density is formed by $z = 7, 6, 5, 4$, respectively.

The inferred steep low-mass-end slope at high redshift indicates that the contribution of low-mass galaxies below our mass limit that we are missing ($M_* < 10^8\ M_\odot$) to the total stellar mass density may be significant if the extension of the Schechter fit is valid at smaller masses than those probed by our sample. Using our best-fit Schechter function parameters, the stellar mass density at $z = 4, 5, 6,$ and 7 would increase by factors of 1.1, 1.1, 1.5, and 1.6, respectively, if the low-mass end of the integral were $\log(M_*/M_\odot) = 6$.

## 7. DISCUSSION

### 7.1. Physical Implications

Figure 9 shows the halo mass functions determined by volume-averaging the *Bolshoi* snapshot mass functions (Behroozi et al. 2013) over the same redshift ranges as those defining our galaxy samples. Comparing the shape of the halo mass function to that of our GSMF, we can see that the shapes become more similar with increasing redshift. Specifically, while at $z = 4$ the low-mass-end slope is clearly shallower than the halo mass function, in contrast to some other studies that found a low-mass-end slope scaling closely with the halo mass function (see Section 5.4), the steepening of our observed GSMF at low masses leads to a more similar slope at $z \geq 7$. This implies that whatever the physical cause of the suppression of galaxy formation in low-mass halos is at $z \leq 4$, it gradually becomes less relevant at $z \to 7$.

Our observations cannot constrain the characteristic mass $M^*$ at $z \geq 6$. This may imply that our volume is too small to capture the needed numbers of rare, massive galaxies. However, the absence of a distinct turnover may also imply that the mechanisms suppressing the massive end of the GSMF are less severe at high redshift. Over the past few years, a number of studies have found that the exponential cutoff of the UV luminosity function appears to weaken at $z \geq 7$ (Bouwens et al. 2014; Finkelstein et al. 2015; Bowler et al. 2015). Those observations could have been interpreted in two ways: either as a result of a decreasing efficiency for feedback processes (or other mechanisms of suppression), or as a reduction in the impact of dust extinction. Because the GSMF is not primarily affected by dust attenuation (modulo its impact on sample selection), a similar observation with the GSMF would imply that the reduced amplitude of an exponential cutoff is due to a cause other than dust, potentially less efficient feedback.

The steepening of the low-mass-end slope with increasing redshift we observe has an implication for the differential mass growth of galaxies. Figure 13 shows the evolution of the GSMF with redshift, presenting a steep increase of more than one order of magnitude in number density of high-mass galaxies ($\log(M_*/M_\odot) \sim 10.5$) over the redshift range $4 < z < 7$. On the other hand, for low-mass galaxies ($\log(M_*/M_\odot) \sim 8$), we detect only a mild evolution of a factor of two increase between $z = 7$ and $z = 4$. This suggests that the evolution of the low-mass-end slope between $z = 7$ and $z = 4$ is driven by the buildup of intermediate- and high-mass galaxies relative to low-mass galaxies, while the number density of low-mass galaxies remains nearly constant, showing the seemingly opposite of the "downsizing" seen at $z < 4$. For a galaxy population that forms the star-forming main sequence with a less-than-unity slope (e.g., Salmon et al. 2015), this behavior is in contrast to what is predicted for the stellar mass growth dominated by a pure smooth in-situ star formation: in such a scenario, the specific stellar mass growth rate for low-mass galaxies is larger than for high-mass galaxies; thus, the GSMF is expected to *steepen with time*. Therefore, the observed flattening of the low-mass-end slope with time may indicate that other processes, such as hierarchical merging or low duty cycle in low-mass galaxies, must be important in the physical processes governing the stellar mass buildup of galaxies at high redshifts. A more detailed exploration of the underlying physical processes can be undertaken by linking galaxies to halos at each redshift and constraining the galaxy stellar mass growth history across time. This can be done by combining techniques such as abundance matching and the mass accretion history of halos inferred from dark matter simulations. We will investigate this in a follow-up paper and show how this observed "upsizing" can be explained without invoking any unplausible scenarios.

#### 7.1.1. Comparison with theoretical studies

Figure 14 compares our GSMFs at $4 < z < 8$ with the predictions from galaxy formation models. First, we compare our measurements with three SAMs described in Lu et al. (2014) and briefly summarized in Section 4.3. These SAMs are based on the same halo merger trees extracted from the *Bolshoi N*-body simulations, but each employs different recipes for modeling the baryonic physics and makes different predictions for the observables. One difference is the different implementation for the outflow mass-loading factor due to stellar-driven winds, which is the ratio of mass ejection to the star formation rate. While these SAMs parameterize the outflow mass-loading factor as the same functional form of a power law in halo circular velocity, the values of parameters describing the function—the normalization and the power-law slope ($\beta$)—vary between models. This is because, while the velocity of outflows can be constrained relatively well from observations of blushifted interstellar absorption lines probing the material in outflows, it is hard to place a tight constraint on the mass of the outflowing material directly from observations (e.g., Heck-man et al. 1990; Martin 2005). Lu et al. (2014) compared the observables predicted from these SAMs and found that the low-mass-end slope of the GSMFs has a clear correlation only with the parameters describing the outflow mass-loading factor or the timescale for ejected gas to return to the disk and not with other parameters. This suggests that the low-mass-end slope of the GSMFs may be able to provide an alternative constraint on the physics of outflows and can provide insights into the processes responsible for the deficit of baryons in galaxies relative to the cosmological baryonic fraction.

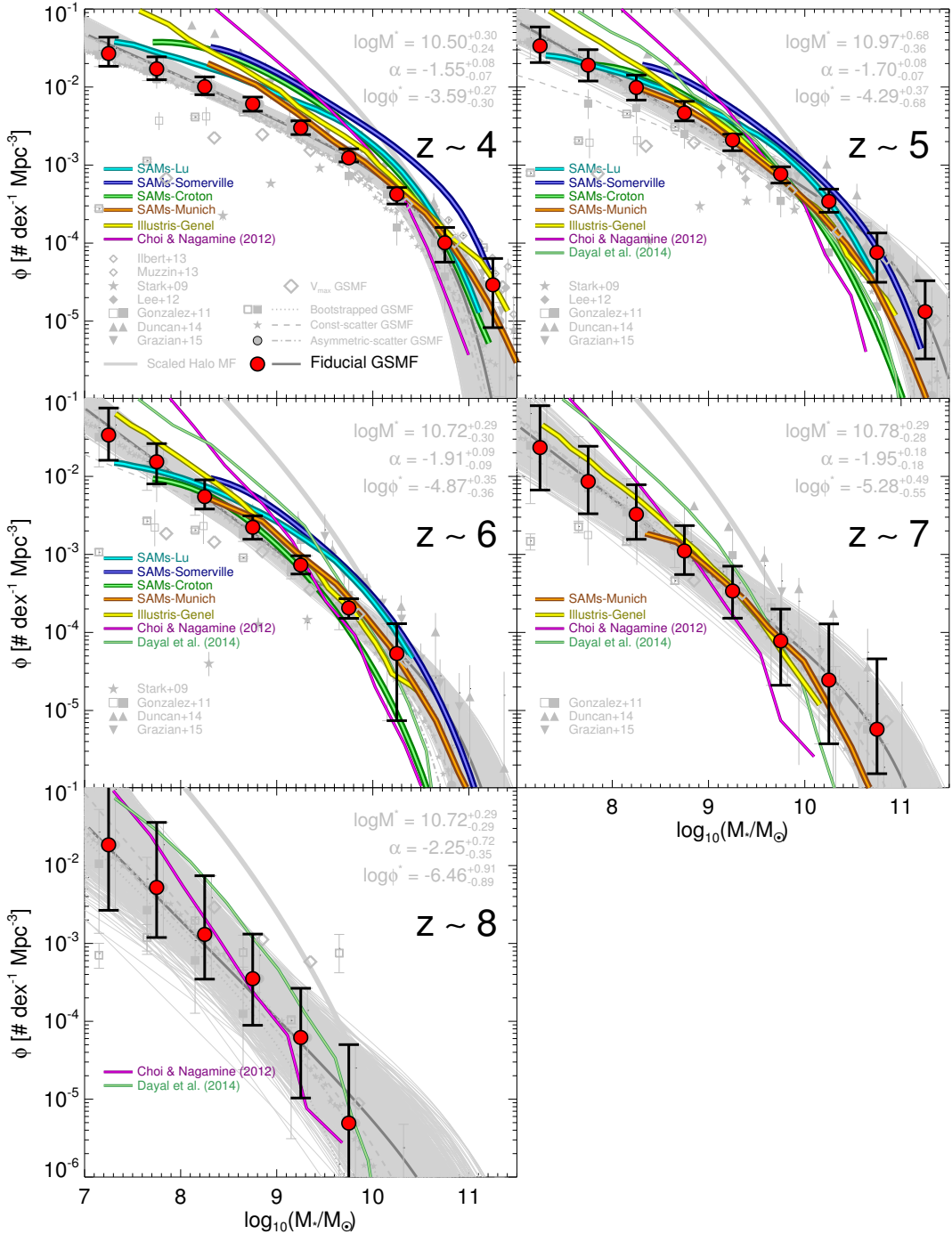

Fig. 14.— Comparison of the observed GSMFs (red circles) and their best-fit Schechter functions (gray solid lines) in this work with theoretical predictions. We show the results from a set of hydrodynamical simulations (the Illustris simulation [Genel et al. 2014], Choi & Nagamine 2012, and Dayal et al. 2014) and SAMs (the Croton model, the Somerville model, the Lu model [described by Lu et al. 2014], and the Munich model [Clay et al. 2015]). The GSMFs predicted from the SAMs are convolved with a lognormal distribution of redshift-dependent standard deviation $\sigma = 0.2$–$0.4$ dex to account for the measurement errors in stellar mass.

Figure 14 compares our GSMFs at $z = 4$–6 with the SAMs.[25] Focusing on the low-mass-end slope, the Lu model, which implements the strongest outflows in low-mass galaxies (i.e., set by a steep dependence of the mass-loading factor on halo circular velocity of $-3.3 < \beta < -9.9$), predicts somewhat shallower low-mass-end slopes of $\alpha = -(1.52 \to 1.61)$ at $z = 4 \to 6$ than observed, while the Croton model with the weakest outflows in low-mass galaxies ($\beta = 0$) has steeper slopes of $\alpha = -(1.68 \to 2.17)$ at $z = 4 \to 6$. The Somerville model with $\beta = -2.25$ displays the most similar low-mass-end slopes to ours ($\alpha = -(1.64 \to 1.93)$ at $z = 4 \to 6$).[26] However, none of the models match the observed GSMFs over the entire redshift range. As noted by Lu et al. (2014), the three models are consistent within the $1\sigma$ confidence level of the Lu model, demonstrating the need for further constraints from observations.

The comparison in Figure 14 also includes the results from cosmological hydrodynamics simulations by Genel et al. (2014), Choi & Nagamine (2012), and Dayal et al. (2014), among which we limit our discussion here to the cosmological smoothed particle hydrodynamics simulation Illustris (Genel et al. 2014), in which energy-driven outflows are implemented ($\beta = -2.0$; Vogelsberger et al. 2013). The simulation with a dark matter mass resolution of $6.26 \times 10^6 \ M_\odot$ was run in a box $\sim (100 \ \mathrm{Mpc})^3$ and was tuned to reproduce the observed GSMF at $z \sim 0$ and the evolution of the cosmic SFRD. Figure 14 shows that the predictions from Illustris are consistent with our observations at all masses at $z = 6$–7 and at $\log(M_*/M_\odot) > 9$ (10) at $z = 5$ ($z = 4$). However, this model increasingly overestimates the abundance of low-mass galaxies with the discrepancy increasing from $z = 6$ to $z = 4$. Thus, while the model is in qualitative agreement with our results in that it predicts a steepening of the low-mass-end slope with increasing redshift, the evolution in the model is milder than observed in that the $z = 4$ low-mass-end slope is much steeper than we observe. The simulation is also known to overpredict the number density of low-mass galaxies at $z \sim 0$ and also at intermediate redshifts $z = 1$–2 (Genel et al. 2014). This suggests that the recipes regulating the stellar mass growth via, e.g., stellar feedbacks in simulations are still overly simplified. The Illustris model does provide a good match at the low-mass end to the observed GSMFs at $z = 4$ by Duncan et al. (2014) and Grazian et al. (2015), but at

$z \geq 5$, the normalization of the GSMF from the model is lower than theirs, in a better agreement with ours. The good agreement in the high- and intermediate-mass range between the model prediction from the Illustris and our observed GSMFs may be attributed to the fact that the model has an extra tuning to match the evolution of the cosmic SFRD in addition to the local GSMF.

Lastly, we compare our GSMFs with the Munich SAM from Clay et al. (2015). In its latest version of the model (Henriques et al. 2015), they tuned their model parameters to reproduce the observed evolution of the GSMFs and passive fractions of galaxies at $z \leq 3$. Among the model parameters, they found that the problem of overproducing the number density of low-mass galaxies at $z > 1$, which has been known to be common to many theoretical models, can be solved by implementing a halomass-dependent timescale for the reincorporation of gas ejected by winds onto the disk. In this new scheme, strong winds and a long timescale for the reincorporation of ejected material in low-mass galaxies delay their growth at high redshift until the ejected gas is finally returned at lower redshift ($z < 2$; Henriques et al. 2015). By applying the same model parameters (except the dust model) at higher redshifts, Clay et al. (2015) predicted the properties of galaxies at $4 \leq z \leq 7$, which are compared with our GSMFs in Figure 14. As shown in the figure, the agreement is remarkable: the model predictions show an excellent agreement with our measurements in the normalization, as well as in both the high- and low-mass-end slopes, except that the model slightly overpredicts the abundance of low-mass galaxies at $z \sim 4$ ($< 0.2$ dex at $\log(M_*/M_\odot) = 8.5$). Especially, the normalization of the model at $z \geq 6$ is lower than that in other observations but is consistent with ours. Understanding what physical process in the model is responsible for the lower normalization at $z \geq 6$ would be interesting to investigate in the future.

### 7.2. Uncertainties

Section 5.4 compared our GSMFs with other estimates from the literature and found considerable discrepancy between different studies. Although one of our two fields used here (GOODS-S) was also used by Duncan et al. (2014) and Grazian et al. (2015), and thus cosmic variance is not likely to explain the discrepancy, none of the previous GSMF studies includes cosmic variance in their error bars. We explore the uncertainties in our GSMFs due to cosmic variance to see whether it can resolve the observed discrepancies.

### 7.2.1. Cosmic Variance

Deep surveys, probing small volumes, are subject to an uncertainty in the number density of galaxies due to underlying large-scale density fluctuations (in addition to the general Poisson uncertainty in the counting of objects). This fractional variance in number density is referred to as "cosmic variance", given as the product of the dark matter cosmic variance ($= f(z)$) and galaxy bias ($= f(M_*, z)$; the clustering of galaxies relative to dark matter) squared. Cosmic variance can be quantified empirically by comparing GSMFs from well-separated (uncorrelated) multiple fields. Because we have only two independent fields, GOODS-S and GOODS-N, our data set is insufficient to quantify cosmic variance, so we turn to other means to estimate its potential impact on our results.

---

[25] The lack of sufficient time resolution in the simulation hinders a reliable construction of the SAM GSMF at $z = 7$–8. This problem will be solved in the final release of the light cones, and predictions for the $z = 7$–8 GSMFs will be presented in Yung et al. (2016, in preparation).

[26] We parameterize the SAM GSMF at $8.0 < \log(M_*/M_\odot) < 11.5$ with a single Schechter function.

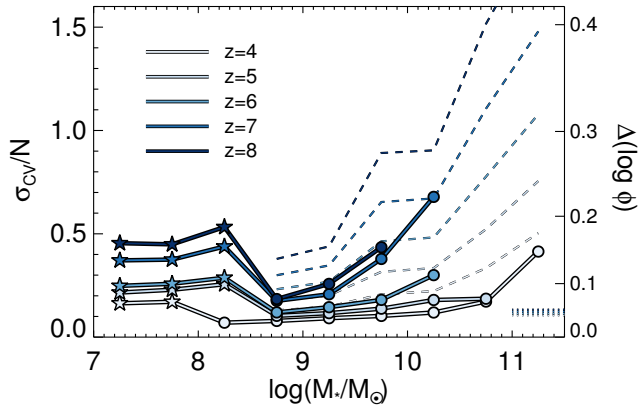

Fig. 15.— Fractional uncertainty on the number density due to cosmic variance, as a function of stellar mass for our survey area (approximated as two $10' \times 16'$ fields for $\log(M_*/M_\odot) > 8.5$ [circles], and a single HUDF-sized field at lower masses [star symbols]). The solid lines represent this quantity estimated from the SAMs (Somerville et al. 2016, in preparation). Dashed lines are values obtained from the tool `getcv` (Moster et al. 2011), and dotted lines (plotted on the lower right side) are from `quickcv` (Newman & Davis 2002; dark matter only, single value at each redshift).

To get mass- and redshift-dependent estimates of cosmic variance for our survey area, we used eight realizations of the GOODS-S field from the SAM of Somerville et al. (2016, in preparation). These SAMs are a reasonable tool for our study, as they have been modified to match the luminosity functions of Finkelstein et al. (2015) used here (the modification was a removal of the birth-cloud component of the dust attenuation; see Finkelstein 2015 for more details). The combined area coverage is about a factor of 40 larger than that of the combined GOODS (N+S) CANDELS fields, allowing us to extract volumes from these catalogs to estimate cosmic variance. We approximate our survey geometry as the two GOODS fields (two $10' \times 16'$ fields), because the two HUDF parallel fields are small ($\sim$10 arcmin$^2$) compared to the GOODS-S field, and galaxies in these fields likely correlate with those in the GOODS-S field given their close proximity.

We cut the full GOODS-S SAM catalog into independent subregions so that their shapes and areas are the same as a single GOODS field ($10' \times 16'$). Then, we calculated the number of galaxies as a function of stellar mass at each redshift ($N(M_*, z)$). The 1$\sigma$ fractional uncertainty on the number density, $\sigma_{\rm cv}/N$, for the GOODS-S field was calculated by bootstrap resampling galaxies in a given stellar mass bin at each redshift. Then, the uncertainty for the total survey volume was calculated by adding the variance for two GOODS-sized fields in quadrature and presented in Figure 15. For $\log(M_*/M_\odot) < 8.0$ (8.5) at $z = 4$ ($z = 5$–8), our galaxies primarily come from the HUDF main field alone. Thus, for those mass bins we estimated the cosmic variance uncertainty for a single HUDF-sized field ($2.'4 \times 2.'4$).

Figure 15 also shows calculations of cosmic variance using the recipe of Moster et al. (2011) with the tool

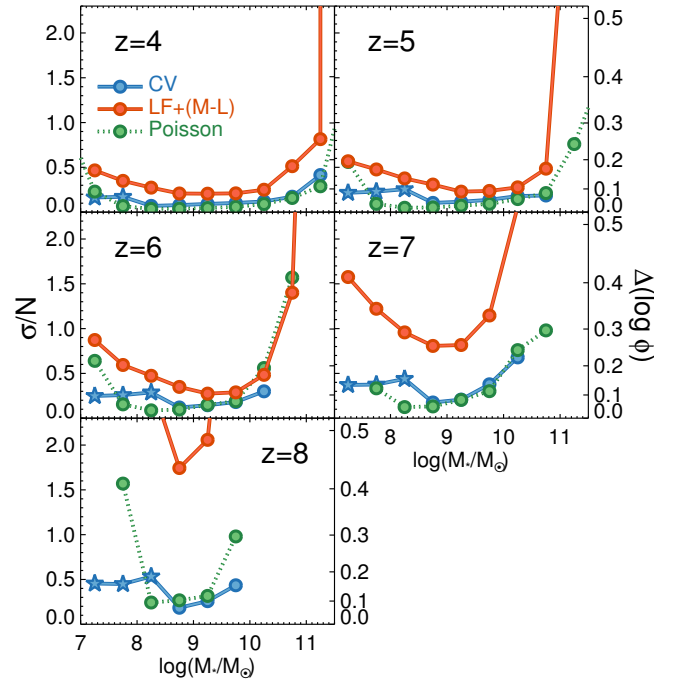

Fig. 16.— Comparison of fractional 1$\sigma$ uncertainties on the number densities of galaxies at a given stellar mass, shown separately for contributions due to cosmic variance, due to the uncertainties of the UV luminosity function and the $M_*$–$M_{\rm UV}$ relation (which includes Poisson uncertainties), and due to Poisson uncertainties alone. The Poisson errors were computed as the half-width of the 68% confidence interval using the recipe of Gehrels (1986).

`getcv` and the method of Newman & Davis (2002) with `quickcv`. The latter gives lower limits on cosmic variance, as it is for dark matter only and therefore does not account for the possible biased clustering of galaxies relative to the dark matter. While the former does include an estimate of galaxy bias as a function of stellar mass, it depends on the extrapolation of the stellar mass–dark matter halo mass relation inferred at lower redshifts ($z < 4$) to higher redshifts, where the stellar mass–dark matter halo mass relation is poorly constrained and the extrapolation may not be valid (Behroozi et al. 2013). Our values for cosmic variance computed from the SAMs are on average a factor of 2 smaller compared to those of Moster et al.

### 7.2.2. *Comparison of Different Sources of Uncertainties*

Figure 16 compares fractional uncertainties in the number density of galaxies in the GSMFs due to cosmic variance to those currently shown in Figure 9, which are due to the combination of uncertainties on the UV luminosity function (which includes Poisson errors) and on the $M_*$–$M_{\rm UV}$ relation.

In general, cosmic variance increases with stellar mass and redshift. Specifically, for a given stellar mass, the larger galaxy bias with increasing redshift leads to larger values for cosmic variance as redshift increases. For a given redshift, as massive galaxies are more clustered, cosmic variance increases toward the massive end of the GSMF. For example, at $z = 4$, cosmic variance reaches up to 40% ($\sim$0.15 dex) for $\log(M_*/M_\odot) = 11.25$, while

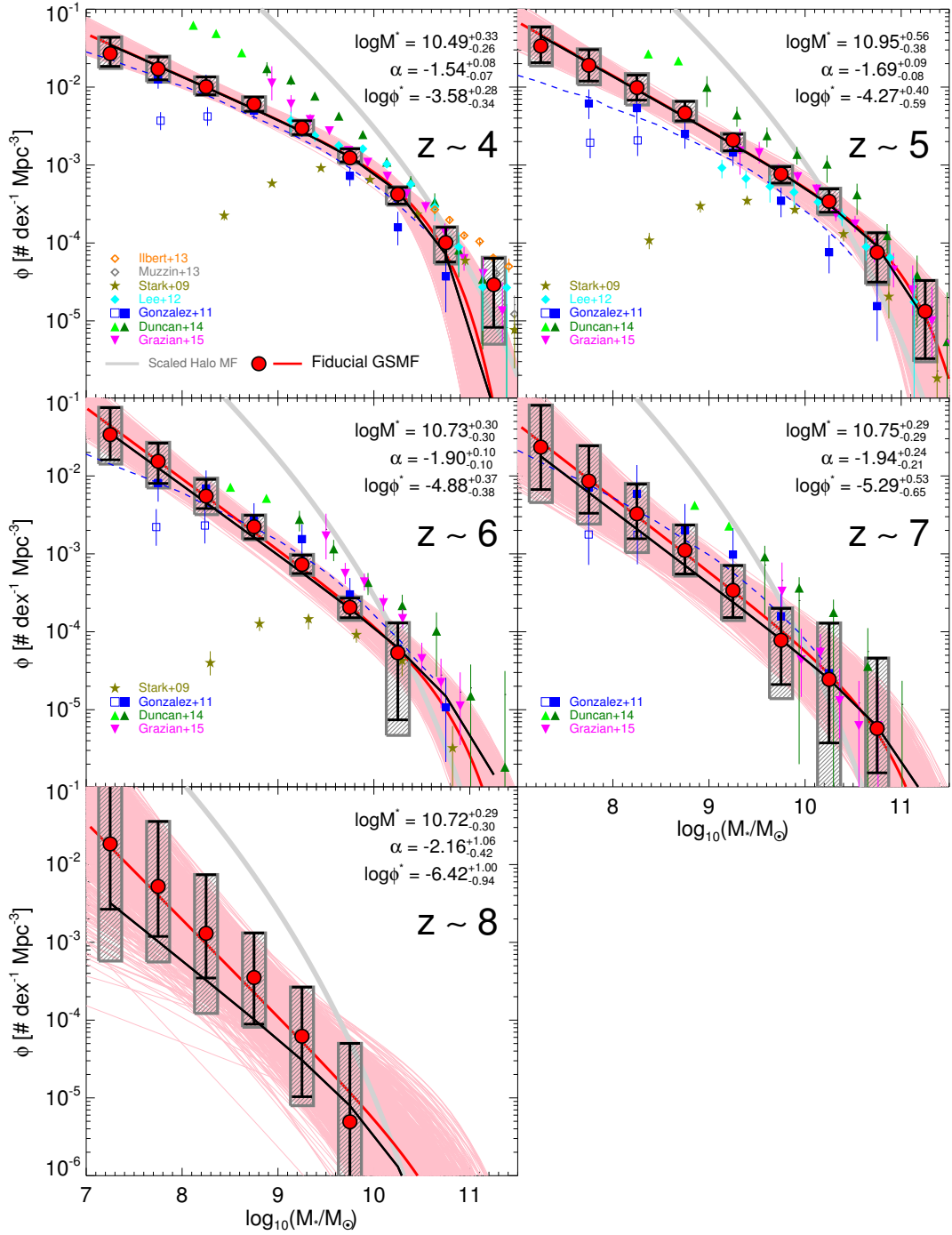

Fig. 17.— Galaxy stellar mass functions at $z = 4$–$8$ (from *upper left* to *lower right*). This figure is the same as Figure 9, only here showing gray shaded boxes which represent the total $1\sigma$ random uncertainties including cosmic variance, the uncertainties of the UV luminosity function, and the uncertainties in the $M_*$–$M_{UV}$ relation. The best-fit Schechter functions derived with the increased uncertainties are shown as the black solid lines, demonstrating that there are negligible differences from the original fit (red solid line).

TABLE 5
Best-fit Schechter function parameters of our fiducial GSMFs refitted with the uncertainties including cosmic variance

| $z$ | $\log M^*$ ($M_\odot$) | $\alpha$ | $\phi^*$ ($10^{-5}$ Mpc$^{-3}$) |
|---|---|---|---|
| 4 | $10.49^{+0.33}_{-0.26}$ | $-1.54^{+0.08}_{-0.07}$ | $26.15^{+24.11}_{-14.08}$ |
| 5 | $10.95^{+0.56}_{-0.38}$ | $-1.69^{+0.09}_{-0.08}$ | $5.43^{+8.22}_{-4.04}$ |
| 6 | $10.73^{+0.30}_{-0.30}$ | $-1.90^{+0.10}_{-0.10}$ | $1.32^{+1.74}_{-0.77}$ |
|  | $(10.97^{+1.35}_{-1.00})$ | $(-1.89^{+0.14}_{-0.11})$ | $(0.75^{+9.18}_{-0.71})$ |
| 7 | $10.75^{+0.29}_{-0.29}$ | $-1.94^{+0.20}_{-0.21}$ | $0.52^{+1.23}_{-0.40}$ |
|  | $(11.08^{+1.24}_{-1.31})$ | $(-1.93^{+0.29}_{-0.21})$ | $(0.21^{+6.20}_{-0.20})$ |
| 8 | $10.72^{+0.29}_{-0.30}$ | $-2.16^{+1.06}_{-0.42}$ | $0.04^{+0.34}_{-0.04}$ |
|  | $(10.26^{+1.42}_{-1.10})$ | $(-1.99^{+1.13}_{-0.56})$ | $(0.15^{+5.78}_{-0.15})$ |

Note. — The quoted best-fit values and $1\sigma$ uncertainties of the Schechter parameters represent the median and the central 68% confidence interval of the marginal posterior distribution of each parameter obtained from our MCMC analysis. The error bars include the uncertainties due to cosmic variance, as well as the uncertainties of the UV luminosity function and the $M_*$–$M_{\rm UV}$ relation. Results in parentheses were derived with a flat prior on $M^*$.

for low-mass galaxies it decreases to $\sim$7% (0.03 dex) for $\log(M_*/M_\odot) = 8.25$. However, as the stellar mass bins at the low-mass end of the GSMFs are dominated by galaxies from the much smaller HUDF, the effect of cosmic variance rises again to $\sim$20% at $z = 4$ and $\sim$50% at $z = 8$.

Due to the small number of galaxies observed at the high- and low-mass ends, cosmic variance for a field comparable to our survey volume is not a dominant source of uncertainty in the number density of galaxies in our GSMFs at the redshifts and stellar masses probed in this study (Figure 16). While the contribution of cosmic variance is comparable to the contribution of Poisson noise to our uncertainty budget, our formal uncertainty, which is the combination of the uncertainty of the UV luminosity function and the $M_*$–$M_{\rm UV}$ relation, is a factor of two larger than cosmic variance. Nonetheless, we examined the impact of the added uncertainty due to cosmic variance on our GSMF. We calculated the total $1\sigma$ uncertainties as the quadrature sum of the two components, $\sigma = (\sigma_{\rm CV}^2 + \sigma_{\rm LF+(M-L)}^2)^{1/2}$. While the increased uncertainties alleviate the tension between different studies to some degree at the massive end, a statistically significant level of discrepancy is still present at all stellar masses and redshifts. This indicates that systematic uncertainties between studies are the likely explanation for the observed differences, as discussed in Section 5.4.

With the increased error bars, we re-fit a Schechter function to our GSMFs with our MCMC analysis (Figure 17). From the fit, we derived the best-fit Schechter parameters listed in Table 5, which show a negligible difference from our fiducial results. In summary, including the impact of cosmic variance on our GSMFs does not change our finding of a steepening of the low-mass-end slope of the GSMFs with increasing reshift.

## 8. CONCLUSIONS

This paper demonstrates the power of combining *HST* with *Spitzer* to explore the stellar mass buildup of galaxies out to $z \sim 8$. Our study is based on a sample of $\sim$4500 galaxies selected via photometric redshifts over $\sim$280 arcmin$^2$ in the GOODS-South and North fields, where deep near-IR and mid-IR imaging data exist from the CANDELS, HUDF, and S-CANDELS surveys.

Our results improve on previous studies in three ways:

- The three-layered depth of CANDELS leads to an increased dynamic range, allowing us to constrain both the high-mass end of the GSMF using data from wide-area, shallow survey fields and the low-mass end using data from deep fields.

- Using the deepest available mid-IR data from the S-CANDELS and the IRAC Ultra Deep Field 2010 surveys with accurate deblending photometry, we have estimated stellar masses more robustly for low-mass galaxies, and subsequently better constrained the low-mass-end slope of the GSMF.

- We have explored and minimized the systematic and random uncertainties inherent in our stellar mass estimation and determination of the $M_*$–$M_{\rm UV}$ relation via simulations using SAMs, which highlight the need for a comprehensive analysis to quantify and minimize the systematics. With the aid of stacking, we can constrain the slope of the $M_*$–$M_{\rm UV}$ relation to within $\pm < 0.1$ of the intrinsic slope at $z \leq 6$ (and also robustly constrain the normalization when fixing the slope at $z = 7$ and 8), lending credence to our GSMFs.

Our main results are summarized as follows:

- Stellar mass and rest-frame UV absolute magnitude are correlated at all redshifts for $\log(M_*/M_\odot) \lesssim 10$. The best-fit $M_*$–$M_{\rm UV}$ relation has a slope marginally steeper than a constant mass-to-light ratio, indicating a higher M/L ratio for massive galaxies than for low-mass galaxies. The slope remains constant over the redshift range $4 < z < 6$, while the normalization shows a weak evolution toward a lower M/L ratio with increasing redshift.

- Taking advantage of the fact that the completeness of the UV luminosity function is relatively well known, we convolved the $M_*$–$M_{\rm UV}$ distribution with a published rest-frame UV luminosity function to derive GSMFs. Our new GSMFs show a clear trend of an evolving low-mass-end slope toward a steeper value with increasing redshift, from $\alpha = -1.55^{+0.08}_{-0.07}$ at $z = 4$ to $\alpha = -1.95^{+0.18}_{-0.18}$ at $z = 7$, providing support for an extension of the trend that is seen at lower redshift but has not been shown (or only marginally hinted at) in previous studies at similar redshifts. Conversely, we find no statistically significant evolution in the characteristic mass, $M^*$, although a larger survey volume will be required to break the degeneracy between $M^*$ and $\alpha$.

- Our GSMFs at $4 < z < 8$ are indicative of differential mass growth of galaxies, where the number

density of massive galaxies increases more rapidly than low-mass galaxies, which is the opposite of the observed behavior at lower redshifts.

- Our estimates of stellar mass density (over $8 < \log(M_*/M_\odot) < 13$) indicate a factor of $10^{+30}_{-2}$ increase between $z = 7$ and $z = 4$, driven mainly by the evolution in the normalization $\phi^*$ of the GSMF, compensated partially by the evolving $\alpha$ toward a shallower slope with decreasing redshift. Comparing our stellar mass density with the time integral of SFRD estimates at similar redshifts shows an excellent agreement at $4 < z < 7$.

While this study provides better constraints on the GSMF at $z = 4$–8, the uncertainties of the GSMFs on both mass ends and at the highest redshift probed in this study are still substantial. Although the advent of *JWST* will make strong advances in this area, the combination of ongoing and planned surveys targeting wide or deep fields will allow us to extend this work to higher and lower masses in the near future. Progress on the high-mass end of the GSMF can also be made with ALMA by placing a more robust constraint on the abundance of dusty star-forming galaxies at high redshift that our current rest-frame UV selection may be missing. On the low-mass end, the Hubble Frontier Field data set, benefit-ting from magnification due to gravitational lensing, enables us to observe galaxies that are intrinsically fainter than the limits of current unlensed surveys. Including the six "blank fields" located near the Hubble Frontier Field clusters, all covered by *Spitzer*/IRAC to the same depth as the S-CANDELS fields, will soon yield more robust constraints on the evolution of the low-mass-end slope of the GSMF.

We thank the anonymous referee for valuable comments that improved this paper. M.S. and S.L.F. acknowledge support from the University of Texas at Austin, the McDonald Observatory, and the NASA Astrophysics and Data Analysis Program through grants NNX13AI50G and NNX15AM02G. S.L. acknowledges the support by the National Research Foundation of Korea (NRF) grant, No. 2008-0060544, funded by the Korea government (MSIP). This work is based in part on observations made with the NASA/ESA *Hubble Space Telescope*, obtained at the Space Telescope Science Institute, which is operated by the Association of Universities for Research in Astronomy, Inc., under NASA contract NAS 5-26555, as well as the *Spitzer Space Telescope*, which is operated by the Jet Propulsion Laboratory, California Institute of Technology, under a contract with NASA.

*Facilities: HST* (ACS, WFC3), *Spitzer* (IRAC)